\vbadness = 10000

\font\sevenrm=cmr7                                                              
                                                             
\font\sevenmi=cmmi7                                                             
\font\sevensy=cmsy7

\font\tenrm=cmr10

\font\tenmi=cmmi10                                                              
\font\tensy=cmsy10                                                              
                                                                                
\font\twelverm=cmr10 scaled\magstep1

\font\twelvemi=cmmi10 scaled\magstep1                                           
\font\twelvesy=cmsy10 scaled\magstep1

\def \twelvepoint{  \textfont0=\twelverm                                        
                    \scriptfont0=\tenrm                                         
                    \scriptscriptfont0=\sevenrm                                 
                    \def\rm {\fam0 \twelverm}                                   
                    \font\it=cmti10 scaled\magstep1                             
                    \font\bf=cmbx10 scaled\magstep1                             
                    \font\sl=cmsl10 scaled\magstep1                             
                    \textfont1=\twelvemi                                        
                    \scriptfont1=\tenmi                                         
                    \scriptscriptfont1=\sevenmi                                 
                    \def\mit {\fam1 }                                           
                    \def\oldstyle {\fam1 \twelvemi}                             
                    \textfont2=\twelvesy                                        
                    \scriptfont2=\tensy                                         
                    \scriptscriptfont2=\sevensy                                 
                    \rm                                                         
                  }                                                             
                                                                                
\parskip=3pt plus 1pt minus 1pt  
\newcount\secno                                                                 
\newcount\subsecno                                                              
\newcount\figno                                                                 
\newcount\eqano                                                                 
\newcount\itemno                                                                
\newcount\iitemno                                                               
\itemno=1                                                                       
\iitemno=1

\def \eno          {\the\eqano\global\advance\eqano by 1}

\def \fno          {\the\figno\global\advance\figno by 1}

\def \iino         {\the\iitemno\global\advance\iitemno by 1 } 
\def \ino          {\the\itemno\global\advance\itemno by 1 }

\def \pno          {\the\pageno}

\def \section    #1{\secno=#1\pageno=1\figno=1\eqano=1 
                    \footline={\hss\tenrm $\sno-\pno$ \hss}} 
\def \sno          {\the\secno} 
\def \ssno         {\the\subsecno} 
 
\def \subsection #1{\subsecno=#1} 
 
\def \page	   {\vfill\eject}


\twelvepoint
\baselineskip=14pt
\parskip=4pt
\smallskipamount=7pt
\medskipamount=14pt
\bigskipamount=28pt  
\hbadness=5000

\leftline{\bf Revised Anatomy of Stars}
\medskip
\leftline{M. Dubin$^1$ and R. K. Soberman$^2$} 

\medskip
\leftline{$^1$ 14720 Silverstone Dr., Silver Spring, MD 20905, 
USA, mdubin@aol.com}
\leftline{$^2$ 2056 Appletree St., Philadelphia, PA 19103, USA, 
rsoberma@mail.sas.upenn.edu}
\bigskip

\noindent{\bf Abstract:} Stars including the Sun, continuously 
accrete near invisible hydrogen dominated agglomerates.  This 
weakly bound ubiquitous baryonic population, `dark matter,' known 
by its gravitational influence on stellar and galactic motion, 
profoundly effects the formation, function and evolution of 
stars.  Measurements, many requiring recent space borne 
instrumentation, provide ample evidence that plasma streams of 
ions, microparticles and macromolecules resulting from the 
disruption of these clusters impact Earth, the planets, the Sun 
and stars.  This sizable mass-energy source contradicts a 
fundamental assumption of the generally accepted nebula collapse 
model of stellar formation.  The visually derived textbook model, 
to which later discoveries (e.g., fusion) were appended, is 
increasingly confounded and contradicted by new observations.  
The discovery of a sizable quantity of radioactive $\rm ^7Be$ (53 
day half-life) in the Earth's upper atmosphere with later tests 
showing it fusion produced, hence coming from the solar outer 
zone, proves the stellar core fusion theory wrong.  Magnetically 
pinched plasma vortices, derived from continuing sporadic capture 
of hydrogen dominated aggregates, impact stars at hundreds of 
kilometers per second, create impulsive local conditions that 
initiate finite nuclear fusion explosions below the photosphere.  
Integral to the aggregate capture process are disks with imbedded 
planets that belt stars.  Giant planets in particular, modulate 
the cluster influx resulting in short term variable fusion rates, 
hence luminosity (e.g., solar cycle).  Recognition of continuing 
accretion of this population, with no assumptions or ad hoc 
physics, explains many stellar phenomena heretofore mysterious, 
e.g., luminosity/wind variation, sunspots, sporadic radial 
magnetic fields, differential rotation, high temperature corona, 
flares, CMEs, etc.  Accepted model phenomenologic explanations 
(where existent) are compared with derivations from continuing 
stellar accretion, e.g., tortuous H-R stellar evolution tracks 
with unobservable (helium flash) kinks contrasted with growth 
along the main sequence curve. 
                                             
 
\page

\leftline{\bf Preamble}

\noindent If you believe astrophysics, as presently taught is a 
valid depiction of the nature of stars, despite ignorance of the 
physical form of $>\! 90\%$ of universal mass; the tacit 
assumption that this dominant mass form's only manifestation is 
in the orbits of stars and galaxies; the continuing series of 
minor theory modifications will someday explain newly observed 
phenomena that puzzle and confound; then we suggest you stop 
reading here.  No argument, even one resting upon a firm 
foundation of observation will change your opinion.  If, however, 
you are troubled by the many inconsistencies and inexplicables 
introduced by recent observations and believe that a fundamental 
change in the way we think about stars is overdue, then open your 
mind and proceed. 

\smallskip
\leftline{\bf 1. Opening}

\noindent Anatomy by definition is associated with biological 
systems.  We used the term in the title to point up 
the similarity between this new view of stars and highly evolved 
biological entities.  Generally accepted stellar theory was built 
upon astronomic observation.  Modifications for subsequent 
detection of unseen mass and radiation were appended to 
hypotheses derived from visual and photographic records.  The 
theory worked, after a fashion, explaining most observations 
(with assumptions added as required), so it persisted, despite 
growing evidence of fundamental flaws.  Hypothetically, stars 
were formed from nebulae collapse because the gravitational 
potential energy was needed to explain the luminosity before 
fusion was recognized.  Fusion was inserted in the core over half 
a century ago after recognition that gravitational energy was 
insufficient for stellar lifetimes.  Today, space borne 
instruments measuring radiation in portions of the spectrum 
unavailable to earlier observers and particle detectors 
unimagined until recently are routinely finding information 
incompatible with or inexplicable by the paradigm.  Although we 
recognize the multitude of nuances and variations; in what 
follows, we combine all into what we call the standard stellar 
model (SSM) as described in current textbooks (e.g., Kippenhahn 
\& Weigert 1994).  SSM is also widely used as the abbreviation 
for the standard solar model (e.g., Bahcall \& Pinsonneault 
1992).  As this is a subset of the standard stellar model we will 
use SSM to refer to both.  It should be clear when we refer to 
the Sun; which we treat as a typical star. 

When a structure increasingly displays cracks, openings and 
displacements upon each new examination, it is frequently the 
foundation at fault.  Examples of the many problems with SSM 
include short period luminosity variations, stellar winds that 
render large stars volatile in comparatively short times, 
multitudes of non thermal surface phenomena, e.g., spots, flares, 
plumes, CMEs, high energy radiation, etc.  For a scientific 
model, the foundation is oft built upon assumptions or theorems.  
SSM rests upon one stated by Russell (Russell et al.\null\ 1927) 
and Vogt (1926) that a star's nature and evolution is determined 
at its birth by a fixed mass and chemical composition.  Nebula 
collapse is thus interpreted as a brief (relative to the stellar 
lifetime) initiating occurrence.  We shall show by citing numerous 
observations that this assumption is wrong; i.e., stellar mass 
accretion is a continuing process.  Significant increasing mass 
over time necessitates major alterations to the accepted model of 
planetary and stellar behavior.  To contrast SSM with continuing 
accretion (SAM for stellar accretion model), we briefly summarize 
both.  

\page
\smallskip
\leftline{\it 1.1 Standard Stellar Model (SSM) Summary}

\hsize=14.5cm
\leftskip 1cm
\noindent A gaseous interstellar nebula, generated by a supernova 
is cooled and compressed (mechanism uncertain) past criticality 
causing rapid gravitational collapse.             

\leftskip 1cm
\noindent Planets, moons, comets and asteroids, condensed 
remnants of the original nebula, orbit the central star(s), 
angular momentum conserved (collisions required to modify 
consistency). 

\leftskip 1cm
\noindent Dust from the nebula, comets and collisions 
(insufficient sources!) form short lived rings and spiral slowly 
into the star due to relativistic (Poynting - Robertson) drag. 

\leftskip 1cm
\noindent Nebula potential energy thermalized in the central 
star(s) initiates fusion in the core(s).  No mechanism known for 
wind or instabilities. 

\hsize=15.5cm
\leftskip 0cm
\noindent In contrast to the foregoing we summarize the 
continuing stellar accretion model (SAM) briefly.  Evidence for 
and consequences of SAM are expanded subsequently. 

\smallskip
\leftline{\it 1.2 Continuing Stellar Accretion (SAM) Summary}

\hsize=14.5cm
\leftskip 1cm
\noindent Cold, weakly bound, largely molecular hydrogen 
aggregates are gravitationally drawn near stars where some are 
trapped to form an orbiting disk.  Sublimation, jetting plus 
disruption enhance the small particle end of the size 
distribution.

\leftskip 1cm
\noindent Interaction with the disk traps more agglomerates.  
Some are directed by gravitation and radiation to impact the 
central star(s) with escape velocity.  Aggregates orbiting in the 
disk continuously grow by accreting some of the infalling mass.  

\leftskip 1cm
\noindent Large masses (planets) within the disk form their own 
smaller disks within which satellites may form or be captured (no 
need for angular momentum consistency).  

\leftskip 1cm
\noindent Aggregate derived hydrogen plasmas enter the stellar 
atmosphere with velocities and forces sufficient to fuse, 
providing mass and energy for radiation and wind. 

\leftskip 1cm
\noindent Stellar wind and radiation interact with infalling 
agglomerates building more complex nuclei, atoms and molecules. 

\hsize=15.5cm
\leftskip 0cm
\smallskip
\leftline{\it 1.3 Observation derived themes}

\noindent In revising stellar anatomy, we develop four 
observation derived themes.  Beginning with continuing accretion 
and evidence thereof, we next discuss the effect of continuing 
accretion upon the nature of stars, with emphasis upon the fusion 
process.  Third we present experimental proof that for the Sun, a 
typical star whose proximity allows greater knowledge, SSM is 
wrong and only SAM can explain recent measurements.  Fourth we 
develop the case against SSM and for SAM.  
  
\smallskip
\leftline{\bf 2. Continuing Stellar Accretion (SAM)}

\noindent Stars will accrete matter gravitationally if it 
penetrates their radiation pressure and stellar wind barrier.  
Clustered or agglomerated matter with small area to mass ratio 
gets through. Intergalactic velocities (Zwicky 1937) and 
galactic rotation (Rubin et al.\null\ 1980) demonstrate that 
abundant mass exists for continuing stellar accretion.  

In cold, field free, intergalactic space, hydrogen with some 
helium plus trace constituents weakly bond by contact
forces.  Collision energy is radiated.  The absence of appreciable 
intergalactic electromagnetic absorption by mass that exerts its 
gravitational influence implies large scale clustering.  
Interstellar space is cold.  Eddington (1926) calculated that, 
with only the energy of stars to warm them, 
the radiative equilibrium temperature of diffuse interstellar 
matter is about 3.18~K.  This is observationally 
confirmed by the cyanogen molecule (CN) that provides a near 3~K 
interstellar space thermometer.  The relative populations of 
the ground and first excited states as calculated from the 
absorption spectrum provides a means of determining the molecular 
temperature.  As it is improbable that the molecule could be 
collisionally excited between stars, the relative excitation 
temperature of 2.3~K $(\rm \lambda = 2.6\,mm)$ requires a 
pervasive millimeter radiation environment to explain the 
measurements.  The presence of interstellar CN, a molecule 
routinely measured in comet spectra, further supports 
cluster/agglomerate existence in and near the galaxy.  That it is 
observed in interstellar rather than intergalactic space is in 
keeping with limited dispersion from the population in the 
comparatively radiation and particulate cluttered galaxy.  

Stellar capture of these weak hydrogen dominated clusters is a 
multibody interactive process.  In gravitational swingby, a few 
will be dispersed by stellar flare and wind encounter, going into 
eccentric stellar orbit.  Poynting - Robertson radiation and wind 
interaction gradually circularize and equatorialize the orbits 
producing a stellar disk.  The disk, extending outward more than 
a light year abets further agglomerate capture.  Infrared and 
submillimeter observations show that at least half the nearby 
stars are surrounded by disks of gas and dust that must be 
continuously supplied (Sargent \& Beckwith 1993).  At the outer 
limits ($\rm R_{disk}\!\sim\! .3\,pc$), where the stellar 
radiation is too weak to sublimate the hydrogen, the clusters 
continue to grow by stellar augmented accretion.  The disk 
perturbs some of the passing aggregates and directs them 
toward the star.  The interaction provides an angular momentum 
bias commensurate with the disk's rotation about the central 
star.  Only those in the far zone (outside the hydrogen 
sublimation range) grow to comet size.  Occasionally, a kilometer 
or larger sized aggregate is perturbed into short period orbit.  
On entering the central zone they display random orbits 
reflecting small perturbations of initially large periastron 
radii.  As with all astronomical distributions, the several order 
of magnitude larger comets are comparatively few in number.  More 
frequently, passing multimeter and smaller agglomerates plus 
perturbed orbiters will be directed toward the star.  Recent 
ultraviolet observations with the Goddard High-Resolution 
Spectograph on the Hubble Space Telescope evidence such bodies 
within the disk about Beta Pictoris falling toward the star at 
several hundred kilometer per second velocities (Lagrange et 
al.\null\ 1996).   

Radiation and wind interaction, particularly at times of sudden 
onset, produce structural stresses that cause jetting, evaporation 
and complete dispersion of agglomerates.  The separated volatile 
dust and gas form anti-stellar tails.  Freed from the agglomerate 
these tail constituents succumb to radiation pressure, slow and 
if far from the star, reverse direction to join the wind.  
Material jetted and volatilized within several stellar radii 
retain sufficient momentum to impact the photosphere with less than 
escape velocity.  Remnants of the original cluster (considerably 
reduced in size and mass), dust, plasma, atoms and  molecules 
impacting at hypervelocity constitute a continuing supply of 
hyperthermal energy. 
 
The disk is an extension of the star's gravitational reach that 
improves cluster capture.  The enhanced capture increases the 
size and density of the disk with age, i.e., symbiosis.  It is 
important to clarify a key point before proceeding.  Readers 
familiar with studies of stellar disks will question our 
statement of continuing growth.  It must be kept in mind, 
however, that continuing stellar accretion means that stars grow.  
Consequently, assumptions of stellar age (founded upon a finite 
fuel supply) are reversed.  To avoid confusion in the minds of 
most readers familiar with the counterintuitive model of bright 
massive blue stars being newborns, whenever we touch upon stellar 
age we shall qualify young with low mass red or infra-red and old 
with massive bright blue adjectives.  Thus the presence of highly 
developed disks about bright blue stars and absence about dim red 
stars is consistent with both SSM and SAM.  Detections of disks 
about main sequence solar type stars such as $\beta$ Pictoris, 
however, poses a SSM dilemma as such disks have theorized lives 
of fractional millions of years after nebula collapse. 

Gravitation and drag within the disk cause agglomerates to be 
directed toward the central body.  Within several AU, central 
radiation guides meter (and smaller) sized clusters.  Sublimation 
produces a `non-gravitational force' (observed with comets) that 
directs the cluster to the radiating source. 

Continuing accretion is widely accepted on binary stars.  The 
mass is hypothesized to originate from the larger star 
beyond its Roche limit.  How this can penetrate the 
smaller star's radiation barrier is rarely, if ever discussed. 
Agglomerates feed both members of binaries.  The interactive path 
causes dispersion, producing a luminous wind between the pair.  
Changing orientation of this interactive path relative to the 
dominant cluster influx direction produces orbital period 
luminosity and wind modulation.  The identification of a 
candidate disk about the binary BD+31\hskip -1pt$^o$643 (Kalas \& 
Jewitt 1997) show both members accreting mass.  A commentary 
(Lissauer 1997) points out the SSM discrepancy if confirmed. 

Stellar winds are common to all stars.  The larger the influx and 
luminosity, the greater will be the stellar wind. The anomaly of 
luminous O and B stars evaporating in relatively short times 
(Thomas 1993) vanishes.  The feedback between the wind and the 
continuing influx is complex.  Only a few aspects are treated 
below. 

Larger bodies with appreciable gravitational contraction are 
occasionally trapped or, in the enhanced accretion environment, 
grow within the disk.  At the extreme, these form planets that 
may orbit relatively close, some gravitationally retaining 
hydrogen and helium.  The planets continue to accrete a portion 
of the incoming flux and gravitationally direct part to the 
central star.  Because of the disk rotation bias, the planets 
will, in time, develop commensurate near circular orbits in the 
equatorial plane of the star.  In contrast to the nebula collapse 
hypothesis, the rotational and orbital angular momentum of the 
planets need not conform to that of the star and disk.  

The nebular collapse angular momentum problem has vexed SSM 
theorists for many years.  Like a skater bringing outstretched 
arms inward, the central star in a collapsing nebula should spin 
rapidly.  That few stars exhibit high spin rates raises the 
question of what becomes of the angular momentum.  Complex 
magnetohydrodynamic hypotheses are invoked to link the disk to 
jets that erupt from T Tauri stars to explain the missing SSM 
angular momentum (Ouyed et al.\null\ 1997; Ray et al.\null\ 
1997).        

The physics of continuing accretion on a lesser scale, produces 
rings, moons and disks about massive planets.  Inner moons and 
rings (disks) will be equatorial and conform to the angular 
momentum of the planet.  Outer moons, likely captured more 
recently, may orbit and spin different from inner bodies. 

Unlike SSM where disks, planets and moons are remnants of 
nebula collapse (despite a physical demand for continuous disk 
supply), serving no purpose in the stellar function, continuing 
accretion promotes the existence of planets about mature stars.  
Subsequent, we describe how many stellar luminosity cycles 
evidence the presence of orbiting planets.  The disk with 
imbedded planets extends the gravitational reach of stars 
increasing cluster influx.  Once inside the disk, planets help direct 
a fraction of the trapped population to the star.  This is 
seen in the influence of Jupiter changing long to short 
period comets.  

Accretion increases the mass of the planets and produces lesser 
disks about them.  As on the stellar scale, SAM poses no 
requirement for consistent angular momentum, a major SSM dilemma.  
Continuing growth causes one planet to eventually become a 
companion binary star.  In the next section we treat the onset of 
fusion that initiates stellar status, showing how this occurs 
during continuing accretion.  Planetary accretion rates vary not 
only with gravity but also the electromagnetic environment, 
producing differing growth, disk, ring and moon structure.  Thus, 
a binary need not evolve from the largest planet. 

As the mass of the star(s) and satellites increase, so too may their 
separation.  In time, planets drift away from evolved stars.  If 
two massive planets exist in proximity about a lone star, one may 
be ejected (Rasio \& Ford 1996).  The interstellar planet (brown 
dwarf) may later be captured by another star or become a solitary 
protostar.   With continuing accretion, stars begin life as 
infra-red low temperature masses that grow in time.  In contrast, 
the counter intuitive SSM has young stars as giant hot blue 
objects.  Stellar disks, now observed with increasing frequency, 
rather than interstellar nebulae may now be properly called 
stellar incubators.  The symbiosis between the star and its disk 
is like that between a spider and its web.  The interaction 
displays a complexity reminiscent of highly evolved biological 
systems. 

Stellar luminosity cycles are observed on many stars.  On 
comparatively dim cool stars they are large relative to the one 
part in a thousand solar cycle.  Short period luminosity 
variation poses a major dilemma for SSM where fusion is invariant 
over tens of mega- and gigayears. 

The population of aggregates must be adequate to provide the 
accretion rate required to fuel the stellar fusion process. The 
observations of Zwicky (1937) and Rubin et al.\null\ (1980) show 
that there is more than adequate mass.  A crude estimate 
indicates the local interstellar population density (typical?) is 
consistent with measurement.  About $\rm 6[10]^{12}\,kg/s$ 
impacts the sun (see Sect.~2.5).  If about 10\% of the trapped 
mass arrives at the sun, this means \hbox{$\rm\sim\! 10^{-
9}\,M_{\odot}/yr$} is integrated into the solar disk.  With an 
order of magnitude disk interaction cross section of a square 
light year \hbox{($\rm \sim\! 1\,LY^2$)} and a relative velocity 
of \hbox{$\rm\sim\! 10^{-4}\,LY/yr$}, the local density is  
(order of magnitude) \hbox{$\rm \sim\! 10^{-5}\,M_{\odot}/LY^3$} 
\hbox{($\rm\sim\! 10^{-23}\,kg/m^3$)}.  This solar accretion 
estimated value is size related.  Aggregates of initial size 
\hbox{$\rm \ll\! 10\,m$} cannot reach the sun, but are part of a 
population with an exponential size distribution steeper 
than -3.   

The solar system exemplifies SAM physics.  The gas giant planets 
modulate and direct agglomerates falling to the Sun.  The 
influence of Jupiter on comet orbits is well documented.  The 
accretion onto these planets produces exospheric super-rotation 
and a continuing energy supply that exceeds sunlight.  The influx 
interactions are most evident on the anti-solar hemisphere but 
extend around the planet.  As the influx produces a weak plasma, 
magnetic interaction causes symmetry about the planetary magnetic 
equator and a banded upper cloud structure. 
 
The encounter of comet Shoemaker-Levy on Jupiter caused x-ray 
aurora showing how hypervelocity impact can produce high energy 
radiation.  The interaction of the solar wind with comets has 
been construed to power x-radiation now observed with space borne 
instruments (Lisse et al.\null\ 1996).  Comet interaction with 
hypervelocity clusters better explains observed bursts of x-radiation.  

The zodiacal cloud is part of the solar disk.  Sunlight scattered 
from dispersing agglomerates was measured from Pioneer 10 and 11 
(Dubin \& Soberman 1991).  Termed cosmoids (a contraction of 
cosmic meteoroid) it was shown that the scattered sunlight from 
dispersing subliming sub-micrometer particles produced all the 
zodiacal light beyond 1~AU and explained its polarization.  If 
the Pioneer 10/11 measurements were the sole evidence for these 
near invisible ($2-3\%$ albedo) clusters, they might be 
dismissed as spurious.  In the following subsections we show that 
measurements extending back nearly half a century form a massive 
body of proof for continuing solar accretion of this 
population.  

\smallskip
\leftline{\it 2.1 Terrestrial influx}

\noindent \"Opik, certain that interstellar meteoroids must 
exist, concluded over half the sporadic meteors were of 
interstellar origin.  Hotly debated (\"Opik 1950) the idea was 
discarded when no meteors with unambiguous interstellar 
velocity could be measured.  An early indication that cosmoids do 
not produce classical meteors is seen in the discovery by 
Ryle and Hewish (1950) of unusual scintillations ($\rm \sim\! 
0.1\%$) of galactic radio sources observed at wavelengths of 3.7 
and 6.7 meters.  Precise analysis indicated sporadic rapid change 
increased electron concentrations in the ionosphere $F$ region 
(150 -- 300~km) with horizontal extent about 5~km occurred only 
at night (maximum likelihood near midnight) with no measurable 
annual modulation.  Their analysis concluded the source of these 
disturbances must have an interstellar origin, entering the 
ionosphere from solar hyperbolic orbit.  Continued effort showed 
that such increased electron density regions causing the radio 
scintillations occurred in interplanetary space and preceded 
solar wind variation (Houminer \& Hewish, 1972; Tappin, Hewish, 
\& Gapper 1983). 

A decade after Ryle and Hewish's (1950) radio scintillation 
discovery, a Stanford University team studying `over the horizon' 
forward scatter radar at 6 to 30~MHz for military surveillance 
found sporadic magnetic field aligned ionization occurring 
in the ionosphere $E$ (90 -- 120~km) and $F$ regions (Peterson et 
al.\null\ 1960).  It was noted that the amplitude of the echoes 
decreased with radio frequency and were more frequently observed 
at higher latitude (Spokane, Washington relative to Stanford, 
California).  Further, it was noted that these field aligned 
ionization patches began and faded rapidly, occurring most 
frequently at night (peak around local midnight).  Typical 
occurrence was at a height of $\rm\sim\! 200\,km$, many scale 
heights ($\rm\sim\! 6\,km$) above the $\rm\sim\! 100\,km$ 
altitude where classic short period meteoroids produce ionization 
trails that create traditional coherent backscatter radar echoes.  

To have escaped identification in meteor studies, cosmoids must 
have several characteristics.  First: inability to produce 
visual/photographic meteors which would have been measured and 
from which orbits would have been deduced.  As with visual 
meteors, nearly every aspect including orbits, of radio meteors 
have been studied for one half century, hence a second 
characteristic: the ionization trail of this class should be 
incapable of returning measurable backscatter echoes at normal 
radar frequencies.  As rocket and satellite borne detectors for 
meteoric particles have been used for one half century, a third 
need be: unlikely to damage artificial spacecraft surfaces.  As 
such a population should reach the exosphere and likely below, 
the fourth would be: interaction with the atmosphere need be 
undetectable with traditional instruments or if noticeable, 
attributable to established environmental phenomena.  We'll show 
by measurements that cosmoids satisfy all four criteria.               

The hyperbolic solar orbit cosmoids that dominate the 
interplanetary meteoroid population by at least two orders of 
magnitude, was only recently discovered (Dubin \& Soberman 
1991) despite prediction and extensive earlier search for their
existence.  Having concluded from the results of the three 
independent Pioneer 10 and 11 interplanetary dust experiments 
that cosmoids were measured, had to be mostly volatile, extremely 
fragile and easily dispersed; we began an extensive search of the 
literature for further evidence of their existence and 
measurement.  Although not recognized as such by the 
investigators, we concluded that cosmoids were observed in the 
inner solar system by the dust detectors on the HELIOS spacecraft 
(Gr\"un et al.\null\ 1980), the Galileo and Ulysses 
spacecraft (Gr\"un et al.\null\ 1992).  Results from the Ulysses 
dust detector during the Jovian encounter confirmed the 
interstellar origin (Gr\"un et al.\null\ 1993) and demonstrated 
the dispersion and redirection produced by that planet. 

Comet-like (but orders of magnitude smaller) loose ensembles of 
submicron grains of frozen volatiles, mainly $\rm H_2,\; H_2O,\; 
CO$ and $\rm CO_2$, interspersed with gas atoms, molecules and 
minor meteoric species, a fraction of the total cosmoid population 
burst or jet in interplanetary space when structural stresses 
induced by solar warming are increased by solar activity beyond 
critical levels.  Data from the Pioneer 10/11 instruments, 
particularly the optical telescopes of the Sisyphus 
asteroid/meteoroid experiment, showed that sunlight scattered 
from the dispersing volatile particles before sublimation 
accounted for all of the zodiacal light brightness beyond 
1~AU.  The polarization of the zodiacal light is explained by the 
submicron size of these particles (Dubin \& Soberman 1991). 

In near radial solar orbit, directed by gravity and 
`non-gravitational' radiation forces that vary inversely with 
size (Brandt \& Chapman 1981), cosmoids approaching the Earth 
encounter the antisolar geotail and magnetotail containing high 
concentrations of energetic electrons.  These produce high 
coulomb and electromagnetic forces that result in dispersion out 
to tens or even hundreds of Earth radii.  Potentials of 
\hbox{10~ke\hskip -1pt V} and larger have been measured on satellites 
passing through the geotail (Garrett 1981).  Electrostatic 
breakup of the weak structure causes progressive subdivision, 
branching like a tree to form a dispersed cloud of gaseous 
molecular clusters and submicron frozen grains before reaching 
the upper atmosphere.  As a result, atmospheric interaction is 
diffused over too large an area to be detected by traditional 
instruments.  In consequence of this magnetotail disruption, 
unlike most low eccentricity solar orbit meteoroids that have 
been exposed to sunlight for extended periods, cosmoids do not 
enter the atmosphere as a solid (albeit of low bulk density for 
cometary debris).  Although the energy per unit mass is larger, 
dispersion over many square kilometers precludes traditional 
visual or photographic meteor trails.  The Gegenschein, light 
diffracted by the atmosphere and backscattered $(\approx 
180^{\circ})$ to the Earth from the antisolar direction by these 
disrupted particles evidence geotail dispersion (Dubin 1986). 

While there were earlier observations of night time ionospheric 
scintillation (Ryle \& Hewish, 1950) and coherent field aligned 
electron trails (Peterson et al.\null\ 1960), it was not until newly 
available high power VLF (2~MHz) radar built for military long 
range communication was directed at meteors that Olsson-Steel and 
Elford (1987) measured echoes from a population that produced 
coherent electron trails at altitudes above 120~km.  They 
concluded this population has an altitude distribution completely 
different from traditional radio meteors measured at much higher 
frequency and yields impossibly low densities ($\rm <\! 10^{-
2}g/cm^3$) when calculated by classical solid meteoroid analysis.  
They further reported it dominates the meteoroid flux by at least 
two orders of magnitude.  The peculiar high altitudes of these 
radar signals, extending above 140~km, means that the meteors 
form ionized trails where the atmospheric density is reduced by a 
factor of 100 or more relative to the 90 -- 100~km classical 
radio meteor height.  The computed low density can be understood 
as what would result from the atmospheric interaction of a 
freshly dispersed cloud of particles.  

The third cosmoid characteristic listed above was, `unlikely to 
damage spacecraft surfaces.'  One of the concerns in the earliest 
days of rocket and satellite studies was the hazard posed by 
hypervelocity meteoroid impact.  Experiments to measure the 
meteoroid flux were among the first placed on high altitude 
rockets and artificial satellites (Dubin 1960).  A controversy 
developed over the level of the terrestrial micrometeoroid influx 
as momentum sensitive acoustic microphones and recoverable rocket 
collections measured a flux about three orders of magnitude 
larger than crater and penetration sensing instruments that were, 
in several instances, carried on the same satellites (e.g., 
Soberman \& Della Lucca 1963).  As hazard was the primary 
concern, the lesser fluxes were adopted for spacecraft design 
(Kessler 1970) while the larger values were attributed to 
instrument artifact.  Subsequent, an ionization sensitive dust 
detector on the highly eccentric earth orbiting satellite HEOS~2 
measured streams of particles that exceeded the penetration flux 
by 2 -- 3 orders of magnitude (Fechtig et al.\null\ 1979).  The 
necessity for protecting one of two similar instruments on the 
HELIOS interplanetary spaceprobes from direct sunlight revealed a 
second meteoroid population in eccentric solar orbit that was 
unable to penetrate the $3.75\,\mu m$ protective film (Gr\"un et 
al.\null\ 1980).   
                                 
The fourth characteristic of this interstellar 
meteoroid population is, `its interaction with the atmosphere 
need be undetectable with traditional instruments or if 
noticeable, attributable to other atmospheric or meteoric 
phenomena.'  Our search for signatures of energetic streams from 
cosmoids interacting with the ionosphere/exosphere centered on 
trace meteoric constituents.  While not as dramatic as the real 
time measurement of dispersed cosmoid particle streams by radar 
(Olsson-Steel \& Elford 1987) there exist numerous terrestrial 
and satellite observations of cosmoid stream phenomena.  
Individually they are not easily interpreted, but together form a 
pattern providing further evidence of the atmospheric cosmoid 
meteor signature.  Discussion of all or even most of the 
phenomena and their interrelations is beyond the scope of this 
paper.  Rather, mentioned below are upper atmospheric phenomena 
(some involving meteoric trace constituents) discovered in the 
last 15 years with newly available ground and satellite 
instruments that defy explanation by classic short period 
meteors.  Detailed treatments of these and more are subjects of 
future communications. 

Using lidar, von Zahn and Hansen (1988) found thin neutral sodium 
layers that appear suddenly (within minutes) at about 95~km above 
high latitudes, extending horizontally for tens to hundreds of 
kilometers.  Traditional meteors produce diagonal trails that 
extend downward for many kilometers.  A mechanism for converting 
neutral atoms from many such trails rapidly into a horizontally 
extended vertically thin layer is unknown.  To generate sudden 
high altitude sodium layers, Dubin (1989) suggested a mechanism 
from cosmoids, electrostatically transformed by capturing 
energetic electrons in the geotail, to a cloud of near uniformly 
sized molecular clusters.  As the stream is of one velocity, it 
is stopped by the atmosphere in a very narrow altitude range, 
less than a scale height ($\rm\ll\! 6\,km$).  Vaporization, 
sputtering or photolysis of molecular clusters releases atomic 
sodium.  As the cloud of molecular clusters is weakly charged, 
the influx is modified by the Earth's magnetic field, deflecting 
much of the mid latitude influx to polar regions.  The variation 
with latitude of the upper atmosphere sodium, minima at mid 
latitudes, implies that the cosmoid component of the meteoroid 
population dominates the distribution, and also implies early 
dispersion. 

Tuned to an iron resonance line, sudden appearing high altitude 
sporadic neutral iron layers were discovered with lidar by Bills 
and Gardner (1990) and Kane, Mui and Gardner (1992).  These have 
similar spatial characteristics to the lidar found sodium, but 
are measured at a middle latitude.  Iron yields a much greater 
meteoric signature than sodium.  Grebowsky and Pharo (1985) 
using ion spectrometers on satellites have measured iron ions 
above 140~km extending to 500~km altitude.  This is far 
above the 100~km height of classic meteor ablation.  Hypothesized 
upward electromagnetic transport fails to explain latitude and 
local time distribution.  Observed in resonance fluorescence from 
satellites and morphologically mapped from the Space Shuttle, 
Mende, Swenson and Miller (1985) found magnesium ions extending 
to 500~km.  These recent discoveries of obvious meteoritic atomic 
and ionic constituents resist explanation by classical short 
period meteors because altitudes, latitudes, geometry, magnetic 
field alignment and local times of measurement are inconsistent 
with that source.  Rather, these phenomena may be understood and 
explained as resulting from the atmospheric interaction of 
energetic plasma streams produced by the dispersion of cosmoids. 

One upper atmosphere visual phenomenon, noctilucent clouds 
resisted numerous attempts to be described in terrestrial terms 
(Soberman 1963).  Since the late nineteenth century it was known 
that these tenuous clouds, appearing sporadically in the northern 
twilight arch above the summer arctic occur about 80~km high.  
They forward scatter sunlight hundreds of kilometers southward to 
the surface long after it is in shadow.  The predominant 
water/ice cloud droplets would be dissociated by solar UV if 
transported from below.  The recognition of this mostly volatile 
population now permits association of the extraterrestrial water 
with the micrometeoric nuclei on which rocket studies showed they 
condensed (Hemenway, Soberman, \& Witt 1963).  

Ecklund and Balsley (1981) detected coherent radar echoes that 
peaked at about 85~km, extending for horizontal distances larger 
than 30~km over northern Alaska during the summer; altitude and 
distribution in good agreement with noctilucent cloud occurrence.  
Measured ionization at that altitude cannot account for coherent 
radar echoes at the frequencies used.  While the above and other 
cosmoid associated manifestations (e.g., other electromagnetic 
emissions) will receive extended treatment in the future, they 
are mentioned here to establish, beyond question, the terrestrial 
influx of this population. 

\smallskip
\leftline{\it 2.2 Continuing accretion on the giant planets}

\noindent The giant planets with their own disks of rings and 
moons evidence continuing accretion.  Rings that lose 
mass to the planet require continuous resupply (Esposito et 
al.\null\ 1984).  Clues to the mechanism were in the radial 
spokes first observed by the Voyager cameras (Smith et al.\null\ 
1881, 1982) that appear randomly in the shadowed portion of 
Saturn's rings.  The interaction of narrow elongated 
hypervelocity streams of dispersed macromolecules and 
microparticles with the orbiting ring particles produce radial 
spokes that diffuse as the collisional disturbance is dissipated. 
The very intense short radio bursts detected on Voyager 2 by the Plasma 
Wave (Gurnett et al.\null\ 1989) and Planetary Radio Astronomy 
Experiments (Warwick et al.\null\ 1989) near Neptune likely were 
similarly generated by such interactions. 

Dispersion and a strong gravitational dust concentration produced by 
Jupiter were observed during Ulysses' planetary encounter. 
Velocity measurements established that the interacting cosmoids 
were in hyperbolic solar orbit hence interstellar (Gr\"un et 
al.\null\ 1993). 

Stronger evidence for the continuing accretion of cosmoids is 
found in the atmospheric high temperatures, super-rotation and 
banded cloud patterns seen on the giant planets (for which no 
satisfactory explanation exists in SSM).  Spectra of the hydrogen 
Lyman~$\alpha$ line in Jupiter's upper atmosphere show supersonic 
velocities from several to tens of kilometers per second (Emerich 
et al.\null\ 1996).  The prograde bias resulting from interaction 
with the disk, solar wind and more remote planets produces a 
torque in the momentum derived from cosmoid accretion that drives 
the exospheric super-rotation.  The process (fluid shear) is 
treated subsequently in Sect. 2.5 where a solar consequence is 
explained.  Jupiter's supersonic winds and hydrogen bulge, for 
example, derive from the infall mass velocity components, i.e., 
solar escape, Jovian orbital and escape.  The dispersed streams 
impacting on a rotating atmosphere produce longitude spread while 
the electromagnetic interaction of the plasma with the bipolar 
magnetic field results in a banded wind (upper cloud) structure 
that displays symmetry about the geomagnetic equator.   

The influx of this matter at a combination of solar plus planet 
escape speeds profoundly effects the energy budget of the 
planets.  This is of great import for the gas giant planets far 
from the Sun.  Jupiter's exosphere, for example, is heated by the 
infalling dispersed plasmas to a temperature of about 1,000 K as 
measured from Voyager (Festou et al.\null\ 1981) and confirmed 
from the Galileo Probe (Seiff et al.\null\ 1997).  This is also 
the mechanism heating the corona to $2[10]^6$ K (discussed 
subsequent).  As the gravitation energy is predominantly inverse 
to solar distance, it is three orders of magnitude less at 
Jupiter where the influx is mostly restricted to the night 
hemisphere yielding an additional factor of 2 decrease relative 
to the coronal temperature.  Evidence of this influx is also seen 
on Venus, particularly in the shadowed hemisphere. 

\smallskip
\leftline{\it 2.3 The zodiacal influx}

\noindent At extreme temperature within several solar radii ($\rm 
R_{\odot}$), particles in low eccentricity slow Poynting -
Robertson spiral orbit should vaporize, creating a predicted
dust free zone. No dust free zone was found near the photosphere 
in more than two decades of searching.  Two measurements made 
during the 1991 solar eclipse add to the numerous IR observations 
showing the zodiacal dust not just extending but increasing to 
the solar disk (Lamy et al.\null\ 1992; Hodapp et al.\null\ 
1992).  Close to the disk, volatile dust could only exist 
briefly, as in near radial hypervelocity influx, before 
sublimating.  While radiation pressure may exceed gravitational 
pull for ablated dust and gas from high velocity comet-like 
agglomerates, inertia produces continued streaming at hundreds of 
kilometers per second (only slightly decelerated, if at all), in 
the brief period required to reach and penetrate the photosphere.  
Hicks, May and Reay (1974) discovered that the solar corona had a 
significant inward radial velocity from measured doppler shifts 
of several Fraunhofer lines.  With improved doppler measurement, 
Fried (1978) confirmed that within 0.7~AU the dust responsible 
for this component of the $F$ corona was streaming into the Sun 
with near escape velocity.  Using a photoelectric radial velocity 
spectrometer that scanned 17 Fraunhofer lines, Beavers et 
al.\null\ (1980) measured two populations, one in prograde orbit 
beyond $\rm 4\,R_{\odot}$ and a second falling into the Sun with 
velocities from about 50 to 250 km/s at about that same solar 
distance.  This infall represents the continuous streaming of 
particles through the corona, continuing mass accretion by the 
Sun from interstellar space. 

\smallskip
\leftline{\it 2.4 The corona}

\noindent The existence of a coronal temperature of two million 
kelvin (2,000,000~K) above a 6,000 K chromosphere has been an 
enigma.  Attempts to transport energy from the solar interior have 
failed.  Bondi, Hoyle and Littleton (1947) showed that collision 
with infalling interstellar hydrogen would provide the required 
coronal heat, but the idea was discarded when it was recognized 
that gas and fine dust would be blown away by the solar wind.  
Aggregates with an area to mass ratio $\rm<\! 10^3\,m^2/kg$ 
falling with near escape velocity ablate gas and dust that heat 
the corona by collision. 

Announced on 11 June 1996 by experimenters using the Ultraviolet 
Coronagraph Spectrometer (UVCS) aboard the Solar and Heliospheric 
Observatory (SOHO) spacecraft was the detection of one hundred 
million kelvin ($\rm 10^8\,K$) oxygen ions in a `quiet Sun' 
coronal hole.  The temperature had not peaked at $\rm 1.9\, 
R_{\odot}$.  In the same solar structure, protons reached a peak 
of $\rm 6[10]^6\,K$ at $\rm 2\, R_{\odot}$ then gradually fell off.  
Separate measures showed the electrons at about one million 
kelvin (Glanz 1996).  Low mass electrons would be the first to 
reach thermal equilibrium in the comparatively cooler coronal 
hole.  The equivalent temperature of ions falling radially into 
the Sun at escape velocity is: 

\noindent $\rm T \;=\; {{2GM_{\odot}m}\over{3kR}} \hfill (1)$

\noindent For oxygen at $\rm 1.9\, R_{\odot}$ this is $\rm\sim\! 
130[10]^6\,K$ and for protons at $\rm 2\, R_{\odot}$ it is $\rm\sim\! 
7.5[10]^6\,K$.  High mass proportional ion temperatures in the 
corona, baffling under SSM, provide additional proof of 
continuous accretion. 

\smallskip
\leftline{\it 2.5 Differential solar rotation}

\noindent When it was discovered that the tangential velocity of 
spiral galaxies, including our own, does not decrease at large 
distance from the center in a Keplerian manner, additional mass 
beyond the perimeter was theorized to explain the anomaly (Rubin 
et al.\null\ 1980).  Similarly, the accretion of interstellar 
agglomerates with a prograde bias derived from disk, planet and 
solar wind interaction provides the torque to drive differential 
solar rotation (akin to planetary exosphere super-rotation 
above.  No satisfactory internal driving mechanism has been 
offered to explain a tangential velocity that increases with 
increasing distance from the rotation axis.  The equatorial 
velocity, $\sim$2~km/s in comparison to the infall velocity 
$\sim$600 km/s indicates the tangential momentum transfer.  Disk 
(including planetary) interactions also cause the influx to 
increase toward the equator.  The measured variation of the 
rotation velocity with latitude over the solar cycle (Woodard \& 
Libbrecht 1993), inconsistent with internal inertial driving 
mechanisms, follows from the disk modulated influx. 

The differential rotation may be used to provide an order of 
magnitude estimate of the accretion rate that cannot be derived 
from current interplanetary measurements.  If we assume a simple 
two dimensional fluid between two parallel surfaces a distance 
$l$ apart, one of which is moving with a tangential velocity 
$u_l$ relative to the other (Fig.~1), then we may describe the 
shear by the relation: 

\noindent $\rm P \;=\; \mu {{du} \over {dl}} \hfill (2)$

\noindent where $P$ is the pressure and $\mu$ the fluid 
viscosity.  If the pressure results from the infalling 
momentum, we can approximate: 

\noindent $\rm {{v_e} \over {A}}{{dm} \over {dt}} \;=\; \mu {{u_l} 
\over {l}} \hfill (3)$

\noindent where we assume that the influx momentum is transferred 
in the depth $l$ and {\it the interior has negligible spin}.  For 
the solar surface viscosity we use $\rm 10^4\,Ns/m^2$ from the 
kinematic viscosity (the viscosity divided by the density) chosen 
by R\"udiger (1989) in his book on differential stellar rotation 
citing an earlier dissertation on the subject (K\"ohler 1969).  
For the photospheric density we use $\rm 10^{-4}\, kg/m^3$.  We 
estimate $l$ to be $\rm 10^7\,m$ on the assumption that the mean 
free path at that depth should be sufficiently small to 
completely stop the infall.  The mass influx required 
to maintain the observed differential rotation against viscous 
drag is order of magnitude \hbox{$\rm \sim\!10^{-6}\,kg \cdot 
m^{-2} \cdot s^{-1}$.} 

An argument used for many years to show that continuous accretion 
could not gravitationally power the Sun (hence the need for 
fusion), is that measurements of the length of the terrestrial 
year would have detected even small increases in the Sun's mass.  
We do not doubt that fusion provides most of the Sun's power, 
but challenge SSM in where it is occurring (see below).  Using the 
order of magnitude mass influx determined above i.e., 
\hbox{$\rm\sim\! 10^{-10}M_{\odot}/year$}, the annual change over 
several millennia would be several seconds or within long term 
record uncertainties.  

In this section we described a number of observations showing the 
Sun, its disk and planets continuously accrete a significant 
amount of interstellar mass in the form of clusters (cosmoids).  
Included were high coronal temperatures, doppler measurements of 
a radial component to the zodiacal cloud, differential solar 
rotation and several planetary and interplanetary observations.  
Further evidence of continuing accretion is presented in 
describing some consequences of electromagnetic behavior and 
stellar encounter.  As above we rely heavily upon measurements of 
our Sun, a typical star. 

\smallskip
\leftline{\bf 3. Stellar Plasma Impact}

\noindent Agglomerated in intergallactic space, super-cooled 
super-conducting randomly rotating cosmoids may have their own 
bipolar fields.  Near a star, interactive jetting produces a 
plasma tail of microparticles, macromolecules, ions and 
electrons.  

As a rule, cosmoids strengthen a magnetic field they encounter.  
This may seem to contradict Lenz's law.  It does not.  Lenz never 
considered other than electromagnetic forces.  The gravitational 
force in the presence of a massive body overrides the repulsive 
electromagnetic force to the infalling plasma. 

In the absence of a magnetic field, an infalling cosmoid will 
create one.  The plasma, like any fluid develops a Coreolis 
vortex of neutrals and ions.  This generates an axial 
magnetic field in which the electrons flow.  The axial magnetic 
field enhances the vortex by pinching the gravitationally 
accelerating current.  The outward extending field acts as a 
magnetic funnel to guide other plasmas in similar orbits.  Thus 
once such a vortex is established it will persist, strengthen
or weaken, depending on the supply of accreting matter.     

Concentrated by stellar gravity, cosmoids like comets 
are modulated and directed toward the central star by the giant 
outer planets (Cowen 1996).  Only as clusters with an 
area-to-mass ratio sufficiently small for gravitational 
attraction to overcome radiation pressure can this material 
approach a star.  For Sun this limit is $\rm<\! 10^3 m^2/kg$. 
Ablation from stellar heating destroys small cosmoids and causes 
the remainder to shrink in size as they approach the Sun.  Comets 
typically ablate more than a meter of surface thickness to 
generate coma and tail during a single perihelion passage.  Sun 
impacting comets (Michels et al.\null\ 1982), too small for 
detection far from the Sun, are frequently observed in 
coronagraphs to enter the solar disk and disappear.  One, 
photographed by the large-angle spectrometric coronagraph (LASCO) 
on the SOHO spacecraft was featured on the front page of the 
February 25, 1997 issue of Eos, newspaper of the American 
Geophysical Union.  Improved space based coronagraphs will 
increase observations of impacting comets as small size 
increases impact likelihood.

Radiation caused jetting, characteristically observed in comets 
(Sekanina \& Larson 1986) produce the so called non gravitational 
forces (Brandt \& Chapman 1981) that vary inverse to the 
size of the body.  Thus, the star's radiation directs approaching 
cosmoids to impact the stellar atmosphere with near radial, near 
escape velocity.  As the cosmoid composition is hydrogen 
dominated and ablation keeps the interior cold (as observed with 
long period, near hyperbolic orbit comets), hydrogen also 
dominates the remnant.  Within several stellar radii inertia 
permits even ablated gas and submicrometer particles to reach the 
surface only modestly slowed by radiation pressure.     

The impacting plasma, squeezed by the magnetic pinch arrives at 
the solar surface with a velocity of about 600 km/s.  For a 
proton this is 2~ke\hskip -1pt V, well above the critical energy 
for the proton-proton (PP) fusion reaction (Fig.~2).  The 
hydrogen encounters the ambient atmosphere with the force of a 
metaphorical sledge hammer but magnetic pinching converts that 
force into a metaphorical pickax with attendant effect creating 
extreme local high pressure.  While thermalization is 
counterproductive and does not occur until much later, we note 
that the equivalent temperature of the pinched monovelocity 
proton beam is \hbox{$\rm 15[10]^6\,K$}, the hypothesized mean 
solar core temperature.  

If the mean PP reaction time was as long as the classically 
computed core fusion value of $\rm 14[10]^9$ years (Clayton 
1968), apart from increasing the surface deuterium and helium 
slightly, it would be extremely difficult to observe effects.  
Rather, this hypervelocity, downward directed, magnetically 
pinched proton dominated beam reacts many orders of magnitude 
faster with stationary ambient protons than calculated from the 
equivalent temperature.  SSM assumes an isotropic, thermal 
velocity distribution, whereas the beam protons are of uniform 
velocity, unidirectional and magnetically confined.  A few beam 
protons fuse with ambient protons to produce a small number of 
\hbox{Me\hskip -1ptV} deuterons which produce 
\hbox{Me\hskip -1ptV}\ $\rm^3He$ nuclei that fuse to form 
\hbox{Me\hskip -1ptV}\ $\rm^4He$ nuclei and \hbox{Me\hskip -1ptV} 
protons (Fig.~2).  For each \hbox{$\rm\sim\! 600\,km/s$} influx 
proton that fuses, we derive one daughter proton moving with $\rm 
v\!>\! 20,000\,km/s$, equivalent temperature 
\hbox{$\rm>\!10^{10}\,K$} plus additional excited protons that 
collided with the daughter helium nuclei.  It is this energetic 
proton multiplication factor of unity less inevitable losses that 
limits the fusion reactions' explosive growth.  At \hbox{Me\hskip 
-1ptV} velocity, equivalent to a temperature that is three orders 
of magnitude greater than the fusion initiation temperature or 
the hypothesized stellar core temperature, the coulomb barrier is 
overcome and the probability (cross section) for protons to fuse 
with ambient protons is increased exponentially.  Bosman-Crispin, 
Fowler and Humblet (1954) compute the exponent to be 4.5 for 
\hbox{$\rm\sim\! 10^7\,K$} decreasing for higher temperatures.  
For interactions subsequent to the initial influx proton beam 
collision, the cross section or probability of fusion should 
increase by at least 12 orders of magnitude with a similar 
decrease in the reaction time, resulting in an explosion of a 
size proportional to the mass of influx protons.  This estimate 
is many orders of magnitude larger than the theoretically 
calculated PP \hbox{Me\hskip -1ptV} cross section and contrasts 
to measured carbon-nitrogen-oxygen (CNO) hydrogen-to-helium 
fusion reaction cross sections, also involving weak beta decays 
at those energies.  As carbon, nitrogen and oxygen are present in 
cosmoids and the Sun but, as discussed below, observation 
indicates that PP is the dominant solar reaction, it appears that 
the PP reaction parameters and nuclear environment requires 
reexamination. 

Another mechanism that may be involved in stellar 
continuing accretion outer zone fusion is muon catalysis (Alvarez 
et al.\null\ 1957).  In the hypervelocity encounter environment 
short lived muons may replace electrons in hydrogen molecules 
enhancing fusion reactions.

Stellar outer zone nuclear fusion ($\rm F_{oz}$) 
is no longer so bizarre as it would have seemed before Terekhov 
et al.\null\ (1993) observed the 2.2~\hbox{Me\hskip -1ptV} gamma 
ray line of the P(n,$\gamma$)D reaction from the GRANAT satellite 
during the solar flare of May 24, 1990.  Derentowicz et al.\null\ 
(1977) demonstrated that fusion could be produced by high 
velocity impact.   Cluster-impact fusion in the laboratory adds 
credence to $\rm F_{oz}$.  Initial reports measured orders of 
magnitude higher rates of deuterium fusion than expected when 
nanometer heavy water ($\rm D_2O$) clusters were modestly 
accelerated to impact titanium deuteride (TiD) targets (Beuhler 
et al.\null\ 1989).  In subsequent literature exchanges even 
critics agreed that fusion occurred.  In question was the 
possibility of contaminant deuterons accelerated to high velocity 
in the apparatus.  After careful separation, it was announced 
that most of the heavy ($\rm [D_2O]_n$) and normal water ice 
cluster ($\rm [H_2O]_n$) fusion observed was due to 
contamination, however factors of 2-4 enhancement over individual 
DD and DP fusion interactions remained, the consequence of oxygen 
and molecular ion momentum exchanges and yet to be determined 
effects (Bae et al.\null\ 1993).   These results show that when 
plasmas containing clusters of nuclei, including oxygen and 
carbon nuclei collide, fusion is more probable than computed from 
single nucleon-nucleon interaction.  It is possible that higher 
velocity ions formed within the clusters during acceleration, an 
inseparable part of the cluster acceleration process, may be 
mistaken for contamination. 

As carbon is present in stellar atmospheres and cosmoids, at the 
increased explosion proton velocity the CNO cycle or bi-cycle 
(for the two possible reaction chains) is likely to occur even 
though the influx protons may have been too slow to initiate 
these reactions.  We do not detail the CNO paths, except to point 
out that these hydrogen to helium fusion chains require an 
initiation temperature about three times higher than PP.  
However, above the starting temperature, the classically 
computed reaction cross section increases much faster than PP so 
that it is supposed to dominate above \hbox{$\rm 18[10]^6 K$.}  
Bosman-Crispin, Fowler and Humblet (1954) compute a CNO 
temperature exponent of 20 for \hbox{$\rm\sim\! 10^7 K$} 
decreasing to 10 for higher temperatures.  Thus, CNO hydrogen 
fusion, proceeding faster, should dominate at 
\hbox{Me\hskip -1ptV} energy when carbon is present.  
Observations of daughter isotopes in stellar atmospheres and 
neutrinos from our Sun indicate the dominance of PP over CNO 
fusion even for higher surface temperature stars.  This may be 
the result of understating the PP reaction rate at \hbox{Me\hskip 
-1ptV} energies. 

Below the photosphere, opacity hides most fusion explosions 
although some are likely associated with chromospheric UV 
``explosive events,'' while larger reactions are likely the 
regularly observed (but SSM enigmatic) bright points (Golub et 
al.\null\ 1974; Howard et al.\null\ 1979); sporadic transient 
isolated x-ray flashes with temperatures exceeding \hbox{$\rm10^6 
K$} \hbox{($\rm\sim\!1500$} are estimated on the Sun at any given 
moment).  From meteor studies, the atmospheric depth at which 
maximum deceleration (interaction) takes place is where the mean 
free path becomes comparable to the size of the decelerating body 
(McKinley 1961).  Thus, the pinched cosmoid beam likely 
penetrates deep \hbox{($\rm \sim\!10^7\,m$)} before exploding.  As 
discussed later, only in the occasional flare may we observe 
fusion occurring above the photosphere.  As the atmosphere has 
comparatively low pressure to great depth below the photosphere 
in most stars, the fusion fireball expands rapidly.  Discussed 
above, the energetic proton multiplication factor is exactly 1, 
meaning that losses limit the chain reaction to a size 
proportional to the number of protons in the initiating cosmoid 
created plasma beam.  Apart from spots (discussed subsequent), 
these sporadic, normally unconnected, explosions are distributed 
randomly, with the opacity and heat capacity of the 
sub-photospheric fusion region averaging the power produced like 
a cloudy sky diffuses sunlight.  

As most energy production occurs thousands of kilometers beneath 
the photosphere, the deeper central regions remain static except 
for the settling of fusion products and small amounts of higher 
density influx material.  Gradually, differentiation separates 
the elements.  For a population~I composition, elements heavier 
than helium, comprising 3-4\% of the mass, occupy the central few 
percent of the radius, helium fills the core to $\sim\! 0.2$ 
radii and hydrogen dominates above.  With no internal fusion and 
the heat source in a surrounding shell, the temperature within 
is near uniform.  The maximum occurs in the outer zone fusion 
shell with a steep negative temperature gradient from 
there to the photosphere.  This is a convective region made 
turbulent by magnetically confined infall plasma streams and 
radiation propelled escaping solar wind ions.  The photospheric 
temperature is a thermal balance between the several million 
kelvin fusion shell below and radiative cooling to the cold 
($\rm 3\,K$) sky.  Its value is maintained by the heat capacity 
of the entire stellar mass.  As the spatial density of cosmoids 
varies, modulated by gravitational interaction with the 
planets, during periods of reduced influx, thermal energy is 
supplied by the stellar interior to maintain the visual and long 
wavelength (bolometric) luminosity.  With large flux variations 
the luminosity of some stars fluctuate dramatically.  During 
active periods the higher energy (UV, x- and gamma ray) stellar 
luminosity increases with the flux of initiating cosmoids and 
resulting fusion.  

Hypothetical SSM core fusion ($\rm F_c$) and SAM outer zone 
fusion ($\rm F_{oz}$) require different PP reaction rates that 
assure mutual exclusivety.  $\rm F_c$ requires that energy be 
released slowly, commensurate with estimated stellar life.  
Measured luminosity and limited hydrogen set a SSM boundary 
condition for the PP reaction time.  The theoretical 
$\rm\sim\!10^{10}$ year value results in a computed 
deuterium/hydrogen ratio of $\rm 10^{-17}$ in star cores, a value 
inconsistent with surface observation by $\rm >\!10^{12}$ and 
recognized as a problem (Clayton 1968).  It is noteworthy that 
the PP fusion reaction time varies approximately proportional to 
the concentration of the daughter deuterons, i.e., a larger local 
D/H ratio increases the speed with which protons fuse.  The 
relatively high amount of deuterium observed in stellar 
atmospheres supports $\rm F_{oz}$.  A second requirement for slow 
reaction prevents the star from exploding when core fusion 
conditions exist.  To meet these requirements, $\rm F_c$ models 
ignore collision and daughter proton velocity (effective 
temperature) that increase effective cross section and reduce 
reaction time.  In contrast, $\rm F_{oz}$ requires explosive 
reaction, constrained by the infalling hydrogen mass.  Particle 
plus radiation efficiency losses reduce the inherent energetic 
proton unity multiplication factor to yield a fireball with limited 
growth potential.  Thus, the stellar PP reaction rate forms the 
critical difference making outer zone and core fusion 
mutually exclusive. 

\smallskip
\leftline{\bf 4. Proof of continuing accretion driven stellar fusion}

\noindent Two independent teams measured $\rm ^7Be$ on the 
leading surfaces of the Long Duration Exposure Facility (LDEF), 
retrieved after orbiting the Earth for 69 months (Fishman et 
al.\null\ 1991).  The concentration necessary to account for the 
measurements was several orders of magnitude greater in orbit 
than in the stratosphere $\rm \sim\! 300\,km$ below.  In the 
troposphere and stratosphere, radioactive $\rm ^7Be$ (53 day 
half-life) is accepted as the consequence of high energy cosmic 
ray spalling of nitrogen and oxygen mostly between 15 and 20 km 
(Arnold \& Al-Salih 1955).  With orbit altitude production 
inconsequential, a fast vertical transport and concentration 
mechanism was sought.  To bolster that hypothesis, pieces of LDEF 
were examined for $\rm ^{10}Be$, a radioisotope with a 1.5 
million year half-life, similar chemistry, spallation production 
and transport likelihood.  Unexpected, the only $\rm ^{10}Be$ 
found was inherent in the aluminum, as about the same was found 
on interior and control surfaces and it did not exhibit the $\rm 
^7Be$ 100 times leading to trailing surface excess (Gregory et 
al.\null\ 1993).  This leads the investigators to conclude that 
cosmic ray spallation is a most improbable source. 

Fusion is the remaining method to produce $\rm ^7Be$ with the Sun 
the sole source consistent with known astrophysics.  Further, the 
half-life only permits the $\rm ^7Be$ to originate near the 
surface.   A SSM derivative, standard solar model (also SSM) 
tables show core fusion $\rm 10^{19}$ short of providing an 
adequate amount for the LDEF measurements (Bahcall 1989).  A 
straightforward computation (Dubin \& Soberman 1996) shows that 
to obtain the \hbox{$\rm 5[10]^9\ ^7Be/m^2$} the two teams found 
on the leading LDEF surfaces (Fishman et al.\null\ 1991), 
requires most fusion occur near the solar surface. 
These seminal beryllium measurements incontrovertibly prove 
accretion driven outer zone fusion unless/until a physically 
reasonable alternative source is found. 

\smallskip
\leftline{\bf 5. Continuing accretion (SAM) versus nebula collapse (SSM) 
argued}

\noindent We begin with an apology for our characterization of 
the standard stellar model (SSM).  We do not demean nor 
trivialize the profound theoretical efforts incorporated.  
Science is however, humanity's best guesses to explain 
\underbar{limited} observations.  It is the wealth of new 
observations, unimagined during SSM's development, that 
facilitated reappraisal of the founding assumptions. 

The generally accepted model theorizes stars born in the collapse 
of a cold dense dust and molecule containing gas nebula.  
Binaries and clusters form from the same nebula simultaneously or 
nearly so. The stellar mass is fixed (Russell - Vogt theorem) and 
invariant.  Stellar winds and binary mass exchanges later 
modified mass invariance and core fusion was appended. 

In the near gravity free nebula environment coherence is a 
lengthy process, even by cosmic standards, and computer simulated 
collapse in (simulated) reasonable times is neigh impossible 
(Foster \& Boss 1996).  Compounding the difficulty, SSM requires 
collapse times vary inversely with the mass of the resultant 
star(s).  Fusion initiation must be inserted by hand as collapse 
computer models are unable to achieve the requisite conditions.  
An obvious problem is that hydrogen fusion must begin in the one 
place (the center) from which gravitational differentiation would 
quickly remove it. 
 
Bethe and Critchfield (1938) chose the stellar core for the 
required temperature and pressure.  That also provided 
long term stability, commensurate with stellar lifetimes.  
Gravitational variation has a time constant of the order of ten 
million years while the hydrogen fusion process is theorized to 
have a time constant of about ten gigayears.  Photons initiated 
in the stellar core take millions of years to reach the surface 
by random walk scatter.  Long term stability is, however, 
inconsistent with stellar luminosity and wind variation, 
occurring in times extending from milliseconds. 

UV observations of stellar winds present an enigma to SSM.  
Inexplicable in isolated hydrostatically stable stars, the 
strongest would cause massive stars to evaporate before they 
exhausted their theorized fixed fuel supply.  Thomas (1993) 
detailed several aspects of the incongruity between SSM and 
recent wind observations.  

Solar measurements with recently developed instruments confound 
SSM theoreticians.  The situation has grown worse in the near
two decades since Parker (1978) wrote ``Indeed, the activity of 
the sun provides so many effects outside the realm of 
conventional laboratory physics that its contemplations is a 
humbling experience for the serious physicist, repeatedly 
demonstrating the incorrect nature of our best ideas and 
explanations.'' 

If the LDEF beryllium measurements cited in the preceding 
section are proof of continuing accretion, why do we deem it 
necessary to proceed?  Experience has taught that a singular 
proof, lacking an explanation consistent with SSM, will be 
rejected.  In upholding the editorial decision to refuse
publishing our conclusion based upon the cited high signal to 
noise measurements, the Editor-in-Chief of the American Physical 
Society wrote ``It is simply not acceptable to focus on one 
empirical set of data, at the cost of jettisoning an entire body 
of experimental and theoretical work'' (Bederson 1996).  All 
arguments (cum citations) showing that the standard model is 
contradicted rather than supported by experiment were ignored.  
The beryllium results were never in question!

The case against SSM and for a revised continuing accretion 
anatomy of stars is founded upon a plethora of measurements.  For 
each set, we present first the SSM explanation (when we are aware 
of one generally accepted).  Thereafter we provide the continuing 
accretion SAM explanation.  The reader should compare as would a 
jurist.  Beginning with three critical observation sets previously 
listed, the order thereafter takes advantage of information 
developed for the reader who may not be familiar with some of the 
diverse specialties we draw upon.  We admit to a biased 
presentation for the prosecution.  However, the informed reader is 
familiar with volumes of arguments for the defense.  In recalling 
the SSM explanations and interpretations it should be kept in mind 
that, with the exception of neutrinos (see below), only the 
surfaces (and above) of stars  are observed.  All else is 
inference and conjecture. 

\hsize=14.5cm
\noindent{\it Stellar luminosity variations} 
\vskip -2pt 
\leftskip 1cm 
\noindent{\bf SSM:} Incompatible with invariant core fusion, 
there have been many attempts to provide energy storage in 
metastable atomic levels and magnetic field twisting.  None have 
proved convincing.  
\vskip -2pt 
\leftskip 1cm 
\noindent{\bf SAM:} With fusion controlled by a variable surface 
accretion rate, luminosity may change on any time scale, subject 
to moderation by energy transport from the interior. 

\leftskip 0cm 
\noindent{\it Stellar winds} 
\vskip -2pt 
\leftskip 1cm 
\noindent{\bf SSM:} No explanation 
consistent with hydrostatic equilibrium exists.  
\vskip -2pt 
\leftskip 1cm 
\noindent{\bf SAM:} Fusion radiation propelled near surface 
plasma plus sublimation from infalling agglomerates fuels the 
wind and explains its variability. 

\leftskip 0cm 
\noindent{\it Missing nebular angular momentum in stars}  
\vskip -2pt 
\leftskip 1cm 
\noindent{\bf SSM:} Complex disk magnetohydrodynamic braking and jets in 
`young' stars (Ouyed et al.\null\ 1997; Ray et al.\null\ 
1997).   
\vskip -2pt 
\leftskip 1cm 
\noindent{\bf SAM:} Problem unique to SSM.

\leftskip 0cm 
\noindent{\it Angular momentum discrepancies in the solar disk}  
\vskip -2pt 
\leftskip 1cm 
\noindent{\bf SSM:} Hypothesized early planetary collisions.  
\vskip -2pt 
\leftskip 1cm 
\noindent{\bf SAM:} Other than an interactive disk/star/planet 
rotation bias, continuing accretion requires no short term 
consistency.  

\leftskip 0cm 
\noindent{\it High temperature giant planet exospheres} 
\vskip -2pt 
\leftskip 1cm 
\noindent{\bf SSM:}  Van Allen particle precipitation, 
ionospheric current heating and gravity wave dissipation 
have been proposed.  
\vskip -2pt 
\leftskip 1cm 
\noindent{\bf SAM:} Accretion heating matches measured 
temperatures (see Sect.~2.2).                    

\leftskip 0cm 
\noindent{\it Planetary exosphere super rotation} 
\vskip -2pt 
\leftskip 1cm 
\noindent{\bf SSM:}  No explanation.  
\vskip -2pt 
\leftskip 1cm 
\noindent{\bf SAM:} Driven by accretion (see Sect.~2).                    

\leftskip 0cm 
\noindent{\it Banded cloud structure on giant planets} 
\vskip -2pt 
\leftskip 1cm 
\noindent{\bf SSM:} Planet rotation plus internal heating explanations sought.  
\vskip -2pt 
\leftskip 1cm 
\noindent{\bf SAM:} Driven by accretion (see Sect.~2).                    

\leftskip 0cm 
\noindent{\it Terrestrial exospheric VLF meteors} 
\vskip -2pt 
\leftskip 1cm 
\noindent{\bf SSM:} No explanation.  
\vskip -2pt 
\leftskip 1cm 
\noindent{\bf SAM:} Accreting cosmoids (see Sect.~2).                    

\leftskip 0cm 
\noindent{\it Sudden appearing meteoric trace constituent layers} 
\vskip -2pt 
\leftskip 1cm 
\noindent{\bf SSM:} No explanation.  
\vskip -2pt 
\leftskip 1cm 
\noindent{\bf SAM:} Produced by dispersed cosmoids (see Sect.~2). 

\leftskip 0cm 
\noindent{\it Noctilucent clouds} 
\vskip -2pt 
\leftskip 1cm 
\noindent{\bf SSM:} A rapid vertical 
water transport mechanism is sought.  
\vskip -2pt 
\leftskip 1cm 
\noindent{\bf SAM:} A result of magnetically directed accretion.   

\leftskip 0cm 
\noindent{\it Gegenschein} 
\vskip -2pt 
\leftskip 1cm
\noindent{\bf SSM:} No explanation.  
\vskip -2pt 
\leftskip 1cm 
\noindent{\bf SAM:} Refracted sunlight backscattered from the 
dispersed cosmoid microparticle geotail (see Sect.~2).               

\hsize=15.5cm
\leftskip 0cm 
\noindent{\it Solar cycle - terrestrial climate link} - 
Climatologists have long been aware of the correlation between 
terrestrial climate and the solar cycle.  Particularly evident in 
tree rings (e.g., Currie 1992), the link has been taken seriously 
only in recent decades because SSM provides no reasonable 
causality.  Solar luminosity varies only in the third significant 
digit, too little to significantly effect global weather.  

\vskip -2pt
\hsize=14.5cm
\leftskip 1cm 
\noindent{\bf SSM:} No explanation.  
\vskip -2pt 
\leftskip 1cm 
\noindent{\bf SAM:} Continuing accretion modulated by the giant 
planets varies $\pm 25\%$ or more, creating the solar cycle.  
With that variation, the terrestrial water influx of cosmoidal 
origin descending through the dry stratosphere produces albedo 
altering cirrus clouds plus other climate modifiers. 

\hsize=15.5cm
\leftskip 0cm 
\noindent{\it Terrestrial aurora precede solar disturbance} -
Silverman (1983) created a perplexing dilemma for theorists when 
he discovered that terrestrial aurora predict solar activity.  

\vskip -2pt
\hsize=14.5cm
\leftskip 1cm
\noindent{\bf SSM:} Aurora are consequences of solar disturbances 
propagated through the solar wind, despite the wind's inability 
to penetrate the magnetopause (Chappell et al.\null\ 1987). 
In light of Silverman's discovery, this explanation violates 
causality. 
\vskip -2pt
\leftskip 1cm 
\noindent{\bf SAM:} The mechanism by which accreted dispersed 
aggregates are magnetically directed to produce the aurora is a 
subject for future papers.  However, it is readily understood 
that enhanced numbers of such aggregates must first transit the 
Earth's orbit before arriving on the Sun to cause a disturbance. 

\leftskip 0cm 
\noindent{\it Rings around planets} 
\vskip -2pt 
\leftskip 1cm 
\noindent{\bf SSM:} Gravity escaping 
debris from meteoric erosion of moons provide 
the necessary continuing particle source.  Despite recognition of 
an interstellar source at Jupiter, Gr\"un et al.\null\ (1996) 
deem it insufficient to explain the measured dust 
concentration.  
\vskip -2pt 
\leftskip 1cm 
\noindent{\bf SAM:} Dispersion of agglomerates provides the 
observed captured rings, accretion and hyperbolic encounter 
microparticle streams.   

\leftskip 0cm 
\noindent{\it X-rays from comets} 
\vskip -2pt 
\leftskip 1cm 
\noindent{\bf SSM:} Solar wind 
explanations have difficulty with cometary x-ray and extreme UV 
emission, particularly bursts of such radiation (Lisse et al.\null\ 1996). 
\vskip -2pt
\leftskip 1cm 
\noindent{\bf SAM:} Collisional excitation of ablated mass from 
hypervelocity clusters provide the requisite energy and explain 
why such radiation is more likely near perihelion. 

\leftskip 0cm 
\noindent{\it Zodiacal matter accretes on Sun} 
\vskip -2pt 
\leftskip 1cm 
\noindent{\bf SSM:} A 
non existent `dust free zone' around the Sun was predicted.  
\vskip -2pt
\leftskip 1cm 
\noindent{\bf SAM:} Several independent experiments have found a 
population falling toward the Sun with near escape speed (see 
Sect.~2). 

\leftskip 0cm 
\noindent{\it High temperature corona} 
\vskip -2pt 
\leftskip 1cm 
\noindent{\bf SSM:} Theoretical 
efforts to heat the corona from below have, without exception, failed.  
\vskip -2pt
\leftskip 1cm 
\noindent{\bf SAM:} The calculations of Bondi et al.\null\ (1947) 
show how sublimated mass from infalling clusters 
(which they did not consider) heat the corona to $\rm 2[10]^6\,K$. 
As shown in Sect.~2, this also explains the SOHO $\rm [10]^8\,K$ oxygen 
measurement (Glanz 1996). 

\leftskip 0cm 
\noindent{\it Differential solar rotation} 
\vskip -2pt 
\leftskip 1cm 
\noindent{\bf SSM:} Physics 
provides no mechanism to internally cause the outer extremes of a 
body to continue to spin faster than the body.  
\vskip -2pt
\leftskip 1cm 
\noindent{\bf SAM:} Continuing accretion provides the required 
external power source. 

\leftskip 0cm 
\noindent{\it Helioseismology} 
\vskip -2pt 
\leftskip 1cm 
\noindent{\bf SSM:} Short period (e.g., 5 min.) 
solar surface acoustic oscillations are attributed to the 
turbulent subsurface layer (Hill et al.\null\ 1996) as resonant 
wave motion can not be sustained far from the power source.  
Long period ($\rm >\!10^7yr$) invariant core fusion, removed 
$\rm \sim\! 1R_{\odot}$ from the surface, is near impossible to 
reconcile with sustained rapid acoustic oscillation.  
\vskip -2pt
\leftskip 1cm 
\noindent{\bf SAM:} Continuing accretion and associated fusion 
results in outer zone turbulence powering acoustic eigenmode 
surface resonances. 

\leftskip 0cm 
\noindent{\it Sunspots} 
\vskip -2pt 
\leftskip 1cm 
\noindent{\bf SSM:} No satisfactory explanation exists.  
\vskip -2pt
\leftskip 1cm 
\noindent{\bf SAM:} In Sect.~3 it was explained how infalling 
plasma vortices create vertical magnetic field funnels that guide  
continuing supplies of accreting plasma.  A large sustained influx 
causes the field lines to strengthen and close, creating a pair 
of spots overlying intense fusion regions.  The cold influx 
explains the cooler temperatures and molecular spectra observed 
(Sandlin et al.\null\ 1986). 

\hsize=15.5cm
\leftskip 0cm 
\noindent{\it Sunspot drift rate} - Libbrecht and Morrow (1991) 
measured the speed with which sun spots travel within the 
photosphere.  They found the small spots move about 2\% faster 
than large spots.  
\vskip -2pt
\hsize=14.5cm
\leftskip 1cm 
\noindent{\bf SSM:} No explanation is compatible with an 
interior driving mechanism.  
\vskip -2pt
\leftskip 1cm 
\noindent{\bf SAM:} With an exterior force applied, Fig.~1 shows 
that the deeper a large vortex extends, the slower it moves 
due to fluid shear drag. 

\hsize=15.5cm
\leftskip 0cm 
\noindent{\it Flares} - Occurring at the apex of magnetic flux 
tubes, they exhibit temperatures $\rm >\!0.5[10]^6\,K$ and are 
accompanied by bursts of gamma radiation indicative of nuclear 
reactions.  The \hbox{2.2 Me\hskip -1ptV} line observed in the 
solar flare of May 24, 1990 showed $^1$H(n,$\gamma$)$^2$H 
occurring (Terekhov et al.\null\ 1993).  The \hbox{0.5 Me\hskip -
1pt V} signature of positron-electron annihilation has been 
seen (Ramaty \& Lingenfelter 1979) and large excesses 
of $\rm^3He$ are associated with solar flare observations 
(Schaeffer \& Zahringer 1962).  
\vskip -2pt
\hsize=14.5cm
\leftskip 1cm 
\noindent{\bf SSM:} Core fusion provides no satisfactory 
explanation for flares.  Hypothetic magnetic reconnection is too 
slow.
\vskip -2pt
\leftskip 1cm 
\noindent{\bf SAM:} A large cosmoid, more common during active 
periods, that cannot be magnetically guided into one of a sunspot 
pair, will occasionally strike the field line apex.  Disrupted by 
the $\rm E \times B$ force, fusion initiated  at the apex, 
interacting with the magnetically associated dual pinched 
reaction zones, provides huge local power sources that may 
persist for hours.  The field lines permit energetic protons plus 
heavier ions to escape. 

\leftskip 0cm 
\noindent{\it Spicules} 
\vskip -2pt 
\leftskip 1cm 
\noindent{\bf SSM:} No clues are offered to the 
origin or nature of spicules that rise 5 to 20 km through the 
chromosphere and may persist for minutes.  
\vskip -2pt
\leftskip 1cm 
\noindent{\bf SAM:} The pinched column of infalling plasma sits 
atop a fusion reaction zone.  Energetic ions propelled back along 
the magnetically confined column interact with the plasma influx 
to a maximum height set by the reaction intensity below.  When 
the influx abates, so too the resultant spicule.   

\hsize=15.5cm
\leftskip 0cm 
\noindent{\it Solar mass ejection} - Also called coronal 
mass ejection (CME), these huge masses \hbox{($\rm 10^{12}\,kg$)} 
depart the sun with velocities $\rm \ge\!400\,km/s$, no 
deceleration, most often near solar maximum (de Jager 1986).  With 
thermal energies $\rm 10^{23}-10^{24} J$, they are accompanied by 
bursts of gamma radiation from the photosphere indicative of 
nuclear reactions.  They are observed on many stars (Drissen 
1992).  Early results from the large-angle spectrometric coronagraph 
(LASCO) on the SOHO spacecraft shows pairing connected by 
quadrant scale magnetic loops (Brueckner 1996).  
\vskip -2pt
\hsize=14.5cm
\leftskip 1cm 
\noindent{\bf SSM:} Hildner (1986) among others dubbed them mysterious. 
\vskip -2pt
\leftskip 1cm 
\noindent{\bf SAM:} Relatively large \hbox{($\rm>\!100\,m$)} and 
consequently sparse cosmoid arrivals producing sizable fusion 
explosions provide the power for these events that may link 
magnetically on a global scale. 

\hsize=15.5cm
\leftskip 0cm 
\noindent{\it Solar neutrinos} - Neutrino detections are unique 
in measuring fusion.  As neutrinos interact extremely weakly with 
nuclei, most will depart the sun independent of the fusion 
locale.  Measurements of four experiments (Homestake, Kamiokande, 
GALLEX and SAGE) detecting different neutrino energies, hence 
fusion reactions (Fig.~2) have created havoc because they 
contradict accepted theory on the following five points: 
\vskip -4pt
\noindent 1) About 30\% to 60\% the number of neutrinos predicted 
are measured. 
\vskip -4pt
\noindent 2) Reported Homestake solar cycle neutrino variation 
(Rowley et al.\null\ 1985), correlated with several solar surface 
related parameters; acoustic (Krauss 1990), magnetic (Oakley et 
al.\null\ 1994), and solar wind (McNutt 1995) violate long term 
invariance. 
\vskip -4pt
\noindent 3) Neutrino flux surges noted in Homestake 
results coinciding with major solar flares (Bazilivskaya et 
al.\null\ 1982; Davis 1987) are forbidden by core fusion.  The 
correlation between a great solar flare and Homestake neutrino 
enhancement was tested in 1991.  Six major flares occurred from 
May~25 to June~15 including the great June~4 flare associated 
with a coronal mass ejection and production of the strongest 
interplanetary shock wave ever recorded that was later detected 
from spacecraft at 34, 35, 48, and 53~AU (Gurnett et al.\null\ 
1993).  It also caused the largest and most persistent (several 
months) signal ever detected by terrestrial cosmic ray neutron 
monitors in 30 years of operation (Webber \& Lockwood 1993).  The 
Homestake exposure (June 1 -- 7) measured $\sim\! 5$ times the flux 
of the preceding and following runs, $>\! 6$ times the long term 
mean and $>\! 2 {{1} \over {2}}$ times the highest measurements 
recorded in $\sim\! 25$ operating years (Davis 1994). 
\vskip -4pt
\noindent 4) Results from two detectors are apparently 
discrepant.  Homestake, believed to measure neutrinos primarily 
from the $\rm ^8B$ reaction (Fig.~2), reports \hbox{$\sim\! 35\%$} 
of prediction, while Kamiokande, that should measure, almost 
exclusively, the same $\rm ^8B$ neutrinos reports \hbox{$\sim\! 
50-60\%$} the predicted value.  
\vskip -4pt
\noindent 5) The various reported neutrino measurements arising 
from the several neutrino producing reactions of the fusion chain 
also produced a `paradox' (Bahcall 1994; Raghavan 1995).  The 
results when compared to model prediction appear to show no 
neutrinos from the $\rm ^7Be$ reaction, but $\sim\! 50\%$ of the 
predicted number from the $\rm ^8B$ reaction.  As $\rm ^8B$ 
results from $\rm ^7Be$ (Fig.~2), measuring neutrinos from the 
daughter reaction while absenting those from the parent is 
paradoxical.   
\vskip -2pt 
\hsize=14.5cm
\leftskip 1cm 
\noindent{\bf SSM:} 1) Predictions from SSM models (e.g., Bahcall 
\& Pinsonneault 1992) exceed solar neutrino measurements by 
165-300\% (Bahcall et al.\null\ 1995).  SSM's apotheosis is 
inherent in the several ad hoc hypotheses generated to explain 
the discrepancies, e.g., the  Mikheyev-Smirnov-Wolfenstein (MSW) 
neutrino flavor change (Mikheyev \& Smirnov 1986) requiring a 
neutrino mass that remains to be reliably measured. 
\vskip -4pt
\noindent 2) All Homestake variations are rejected as 
statistically inconclusive.
\vskip -4pt
\noindent 3) Explained as statistical fluctuation or cosmic ray 
produced (Bahcall 1989).  Attributing the 1991 burst (coinciding  
with the largest recorded solar flare) to statistical variation 
stretches probability to the point where there is as yet no 
comment.    
\vskip -4pt
\noindent 4) No explanation.
\vskip -4pt
\noindent 5) No explanation.
\vskip -2pt
\leftskip 1cm 
\noindent{\bf SAM:} Outer zone fusion driven by varying accretion 
permits resolution of all elements of the solar neutrino puzzle 
consistent with all present measurements (Dubin \& Soberman 
1996), a feat deemed impossible with any modification of stellar 
physics (Bahcall 1996).  

\hsize=15.5cm
\leftskip 0cm 
\noindent{\it Anomalous cosmic rays} - Very low energy 
\hbox{($<\! 10$ Me\hskip -1pt V} per nucleon), they were discovered 
only when cosmic ray detectors traveled beyond the Earth's 
magnetosphere (Hovestadt et al.\null\ 1973; McDonald et al.\null\ 
1974).  Anomalous because their elemental abundances differ 
significantly from galactic and solar cosmic ray populations, 
they increase with solar distance and do not correlate with solar 
proton flux.  Rich in oxygen, nitrogen, helium and neon, Klecker 
et al.\null\ (1977) showed that for those of energy \hbox{$<\! 4$ 
Me\hskip -1pt V} per nucleon the elemental abundances were close 
to those of the solar atmosphere.  
\vskip -2pt
\hsize=14.5cm
\leftskip 1cm 
\noindent{\bf SSM:} No explanation.
\vskip -2pt
\leftskip 1cm 
\noindent{\bf SAM:} The interaction between infalling 
cosmoidal matter, galactic cosmic rays and the solar wind is the 
source of anomalous cosmic rays.  Interaction between the solar 
wind and cosmoids produce an outer layer on the latter that is 
comparatively rich in fusion products.  These include CNO bicycle 
products and heavier elements fused in an environment rich in 
high energy ions.  Neon is favored because it can result from 
several fusion paths. 

\page
\leftskip 0cm 
\noindent{\it Deuterium/hydrogen ratio} 
\vskip -2pt 
\leftskip 1cm 
\noindent{\bf SSM:} As deuterium 
reacts rapidly with hydrogen, it is hypothetically destroyed 
during early nebula collapse (Ezer \& Cameron 1966) and fused 
almost immediately in core fusion.  Slow $\rm F_c$ 
requires ratios of low mass isotopes (particularly deuterium) 
that are extremely disparate from values observed on the surfaces of 
stars, on Earth and in the interstellar medium (Schwarzschild 
1958).  In order to obtain a PP reaction rate slow enough to 
provide observed luminosity throughout a star's life with a fixed 
amount of hydrogen, a $\rm 10^{-17}$ core deuterium/hydrogen 
ratio is calculated; \hbox{$\rm\sim\! 10^{-12}$} the 
$\rm 10^{-4}-10^{-5}$ observed on Earth and in the sky (Clayton 
1968).  To account for the discrepancy, cosmic ray spallation is 
suggested to produce the observed deuterium.   However, that 
mechanism is contradicted by ratios of other low mass isotopes 
(Fowler et al.\null\ 1962a).  
\vskip -2pt
\leftskip 1cm 
\noindent{\bf SAM:} With continuing accretion driven fusion, 
deuterium escapes the fusion region to produce observed D/P 
ratios. 

\leftskip 0cm 
\noindent{\it $^7$\hskip -1ptLi problem} 
\vskip -2pt 
\leftskip 1cm 
\noindent{\bf SSM:} The $\rm^7Li$ observed in 
atmospheric spectra of many main sequence stars including our Sun 
is ascribed to production by cosmic ray spallation.  However, 
the absence of another stable isotope, $\rm^6Li$, that should 
also be created by spallation is referred to as `the lithium 
problem' (B\"ohm-Vitense 1992).  Fowler, Greenstein and Hoyle 
(1962a) illustrate the complex theories required to explain 
low mass isotopic abundances.  Common also in nova spectra 
(Starrfield et al.\null\ 1978) as is the $\rm^7Be$ gamma ray line 
(Leising 1990), inventive models such as slow precursor PP 
fusion (Starrfield 1989) are created to account for their 
presence.  
\vskip -2pt
\leftskip 1cm 
\noindent{\bf SAM:} The daughter of $\rm^7Be$ decay (Fig.~2), the 
presence of \hskip 2pt$\rm^7Li$ in the Sun's atmosphere and those of most 
main sequence stars is a consequence of accretion driven outer 
zone fusion.  Its presence (plus other PP indicators) in novae 
evidence the PP reaction taking place explosively. 

\hsize=15.5cm
\leftskip 0cm 
\noindent{\it 0.5 \hskip -2ptMe\hskip -1ptV gamma ray sky background} - 
Satellite borne telescopes observe a sky background of these 
photons that likely originate from positron-electron 
annihilation.  They are also observed in solar flares (Ramaty \& 
Lingenfelter 1979).   
\vskip -2pt
\hsize=14.5cm
\leftskip 1cm 
\noindent{\bf SSM:} No satisfactory explanation for 
the ubiquity exists.  
\vskip -2pt
\leftskip 1cm 
\noindent{\bf SAM:} Widespread accretion driven PP outer zone 
reactions (Fig.~2) are the likely positron source.   

\leftskip 0cm 
\noindent{\it Dredge-up} 
\vskip -2pt 
\leftskip 1cm 
\noindent{\bf SSM:} To explain observed fusion 
products in population~II red giants and associated nebulae, 
a process called `dredge-up' is hypothesized wherein circulation 
extending from the core to the surface raises them to be ejected 
in the stellar wind.  How hydrogen is retained in the core in the 
presence of such a circulation is ignored.  
\vskip -2pt
\leftskip 1cm 
\noindent{\bf SAM:} Fusion products in stellar atmospheres are 
appropriate. 

\page
\leftskip 0cm 
\noindent{\it H-R main sequence} 
\vskip -2pt 
\leftskip 1cm 
\noindent{\bf SSM:} SSM offers no 
explanation for the shape nor position of the H-R main sequence 
curve.  Hypothesized stellar evolution paths follow complex paths 
on the color luminosity plot, requiring several figures with
unobservable path kinks to show 
how a single mass star ages.  
\vskip -2pt 
\leftskip 1cm 
\noindent{\bf SAM:} The subject of a future paper, the main 
sequence is the continuing accretion stellar evolution path.  A 
star begins as a dim IR source fusing accreted deuterium to 
helium.  The lower inflection on the main sequence occurs when a 
star's mass \hbox{($\rm\sim\! 0.4 M_{\odot}$)} and size 
\hbox{($\rm\sim\! 0.6 R_{\odot}$)} achieve values where the 
infalling magnetic pinched plasma arrives with a velocity of $\rm 
500\,km/s$, initiating outer zone PP fusion.  The upper 
inflection occurs when the mass reaches \hbox{$\rm\sim\! 6.5 
M_{\odot}$} and size \hbox{$\rm\sim\! 4 R_{\odot}$}, resulting in 
an influx velocity of \hbox{$\rm\sim\! 800\,km/s$}.  This will 
begin the triple alpha reaction, fusing helium to carbon (Fowler 
et al.\null\ 1962b).  Spectra for stars above this inflection 
show the presence of strong carbon lines (Underhill 1955; Wilson 
1955).  Departure from the main sequence is the consequence of 
outside (other stars) gravitational interaction. 

\smallskip
\hsize=15.5cm
\leftskip 0cm 
\leftline{\bf 6. Concluding}

\noindent The tally of SSM faults, flaws and inexplicable phenomena 
compared with the observation based continuing accretion model (SAM)
could be expanded many fold; but the point has been made.  In 
Sect.~4 we presented experimental proof, with a large signal 
to noise ratio, obtained by two independent teams, that for the 
Sun, a typical star, the hypothesized nebula collapse -- standard 
stellar (solar) model cannot be reconciled with the short
half-life $\rm ^7Be$ found in the Earth's exosphere.  The only 
explanation consistent with physics is: the beryllium was 
transported by the solar wind from the outer zone of the Sun 
where it was produced by fusion.  
 
Recognition of a near invisible baryonic cluster population 
(cosmoids), the gravitationally noted `dark matter,' for 
which we have selected but a few examples from a plethora of 
observations, allows us with no assumptions, to answer numerous 
riddles posed by recent measurements.  The reader can undoubtedly 
find other examples where this population impacts his/her 
discipline.  In explaining observation, compare, for 
example, continuing accretion stellar evolution along the main 
sequence of the H-R diagram as opposed to the tortuous SSM paths 
requiring several plots for even a single mass star.  
Hypothesized tumultuous kinks in the path are unobservable but 
accepted on faith (e.g., the helium flash).  Counterintuitive 
also is the SSM thesis that the youngest stars are the largest 
and (now measured) the most volatile.  Continuing accretion 
driven outer zone fusion permits understanding of stellar 
luminosity/wind variation and the ubiquity of stellar (planet 
containing) disks that host future stars.  The 
relative simplicity of SAM must appeal to any who can see beyond 
the psychological barrier erected by inculcated theory. 

Committed to SSM, unwilling to relinquish belief in nebula 
collapse, core fusion, etc., supporters cite the myriad of explanations 
underlying that body of hypotheses.  They argue that stars would 
collapse without the heat (radiation pressure) provided by core 
fusion.  Chandrasekhar (1939) found no difficulty in positing 
stellar structure before the invention of core fusion.  He noted 
that the interior would be at several million kelvin if {\it it 
behaved as a perfect gas.}  We use italics to emphasize the 
assumption.  Surrounding a star with a fusion shell satisfies the 
need for internal heat while the density gradient directs 
radiation pressure outward.  

As all stellar measurements with the singular exception of 
neutrinos are surface (or above) observations, interior structure 
is assumed.  The neutrino `problems' have exposed the flaw, with 
numerous (ad hoc physics) assumptions attempting to bridge 
discrepancies.  Outer zone fusion shows all existent neutrino 
experiments yield consistent results (Dubin \& Soberman 1996); 
contradicting Bahcall's (1996) conclusion ``...it does not appear 
possible to reconcile the four operating experiments with any 
modification of stellar physics.''  

Structures may stand, however, long after their foundations have 
deteriorated.  Thus, after providing proof that SSM is flawed and 
fails by comparison to continuing accretion, acceptance may 
continue to be elusive.  Other than those cited above 
there are numerous tests that can distinguish between SSM and 
SAM.  E.g., $\rm^7Be$ should be found in the 
solar wind perhaps by instruments already operating beyond the 
terrestrial magnetosphere; cluster fusion experiments (Beuhler 
et al.\null\ 1989) with $\rm ^1H$ and $\rm ^3He$ in the 
target should yield the \hbox{0.5~Me\hskip -1pt V} gamma ray 
signature of positron annihilation showing the PP reaction 
occurring explosively.  The astute reader will undoubtedly be 
able to suggest more.  We challenge critics to find a measurement 
(as opposed to a SSM derived explanation for same) that is in 
conflict with the foregoing.

\smallskip
\leftline{\bf References}

\def\ref{\hangindent=24pt \hangafter=1 \parindent=0pt}

\ref Alvarez, L.W., Bradner, H., Crawford, F.S., et al. (1957)
  {\it Phys.\null\ Rev.}\null\ {\bf 105}, 1127

\ref Arnold, J.R., Al-Salih, H.A. (1955) {\it Science} {\bf 121}, 451

\ref Bae, Y.K., Beuhler, R.J., Chu, Y.Y., Friedlander, G.,  
  Friedman, L. (1993) {\it Phys.\null\ Rev.\null\ A} {\bf 48}, 4461

\ref Bahcall, J.N. (1989) {\it Neutrino Astrophysics.}  Cambridge 
  Univ.\null\ Press, Cambridge. 

\ref Bahcall, J.N. (1994) {\it Phys.\null\ Lett.\null\ B} {\bf 338}, 276

\ref Bahcall, J.N. (1996) {\it ApJ} {\bf 467}, 475

\ref Bahcall, J.N., Pinsonneault, M.H. (1992) 
  {\it Revs.\null\ Mod.\null\ Phys.}\null\ {\bf 64}, 885

\ref Bahcall, J.N., Lande, K., Lanou, R.E., et al. (1995) 
  {\it Nature} {\bf 375}, 29

\ref Bazilivskaya, G.A., Stozhkov, Yu.I., Charakhch'yan, T. N. 
  (1982) {\it JETP Lett.}\null\ {\bf 35}, 341

\ref Beavers, W.I., Eitter, J.J., Carr, P.H., Cook, B.C.
  (1980) {\it ApJ} {\bf 238}, 349

\ref Bederson, B. (1996) private communication
 
\ref Bethe, H. A., Critchfield, C.L. (1938) {\it Phys.\null\ 
  Rev.}\null\ {\bf 54}, 248

\ref Beuhler, R.J., Friedlander, G., Friedman, L. (1989)
  {\it Phys.\null\ Rev.\null\ Lett.}\null\ {\bf 63}, 1292 

\ref Bills, R.E., Gardner, C.S. (1990) {\it Geophys.\null\ Res.\null\ 
  Lett.}\null\ {\bf 17}, 143

\ref B\"ohm-Vitense, E. (1992) {\it Introduction to Stellar 
  Astrophysics, Vol. 3 Stellar Structure and Evolution.} Cambridge
  Univ.\null\ Press, Cambridge. p.\null\ 84

\ref Bondi, H., Hoyle, F., Lyttleton, R.A. (1947)
  {\it MNRAS} {\bf 107}, 184

\ref Bosman-Crispin, D., Fowler, W., Humblet, J. (1954) 
  {\it Bull.\null\ Soc.\null\ R.\null\ Sci.\null\ Li\`ege} {\bf 9-10}, 327

\ref Brandt, J.C., Chapman, R.D. (1981) {\it Introduction to 
  Comets.} Cambridge Univ.\null\ Press, Cambridge.  p.\null\ 69 

\ref Brueckner, G.E. (1996) {\it Eos, Trans.\null\ AGU,
  77 - Spr.\null\ Mtg.\null\ Suppl.}\null\ S204

\ref Carrie, R.G. (1992) {\it Ann.\null\ Geophys.\null\ Atmos.\null\ 
  Hydro.\null\ Space Sci.}\null\ {\bf 10}, 241

\ref Chandrasekhar, S. (1939) {\it An Introduction on the Study of 
  Stellar Structure.} Univ.\null\ of Chicago Press, Chicago.

\ref Chappell, C.R., Moore, T.E., Waite Jr., J.H., 
  (1987) {\it J.\null\ Geophys.\null\ Res.}\null\ {\bf 92}, 5896

\ref Clayton, D.D. (1968) {\it Principles of Stellar Evolution.}               
  McGraw-Hill, New York.

\ref Cowen, R. (1996) {\it Sci.\null\ News} {\bf 150}, 60

\ref Davis Jr., R. (1987) Report on the Status of Solar Neutrino 
  Experiments. In: Arafune, J. (ed.) {\it Proc.\null\ of Seventh 
  Workshop on Grand Unification, ICOBAN'86 Toyama, Japan.}  
  World Sci.\null\ Press, Singapore. p.\null\ 237 

\ref Davis Jr., R. (1994) {\it Prog.\null\ Part.\null\ Nucl.\null\ 
  Phys.}\null\ {\bf 32}, 13

\ref Derentowicz, H., Kaliski, S., Wolski, J., Ziokolski, Z. 
  (1977) {\it Bull.\null\ Acad.\null\ Polanaise Sci., ser.\null\ 
  sci.\null\ techniques} {\bf 25}, 135

\ref Drissen, L. (1992) {\it PASPC} {\bf 22}, 3

\ref Dubin, M. (1960) {\it Space Res.}\null\ {\bf 1}, 1042

\ref Dubin, M. (1986) {\it Eos, Trans.\null\ AGU} {\bf 67}, 1076 

\ref Dubin, M. (1989) {\it Eos, Trans.\null\ AGU} {\bf 70}, 1243

\ref Dubin, M., Soberman, R.K. (1991) {\it Planet.\null\ Space 
  Sci.}\null\ {\bf 39}, 1573 

\ref Dubin, M., Soberman, R.K. (1996) 
  http://xxx.lanl.gov/archive/astro-ph/9604074
 
\ref Ecklund, W.L., Balsley, B.B. (1981) 
  {\it J.\null\ Geophys.\null\ Res.}\null\ {\bf 86}, 7775

\ref Eddington, A. S. (1926) {\it The Internal Constitution of the 
  Stars.} Dover, New York. p.\null\ 371           

\ref Emerich, C., Ben Jaffel, L., Clarke, J.T., et al. (1996)
  {\it Science} {\bf 273}, 1085

\ref Esposito, L.W., Cuzzi, J.N., Holberg, J.B., et al. (1984) 
  Saturn's Rings: Structure, Dynamics and Particle Properties.
  In: Gehrels, T., Mathews, M.S. (eds.) {\it Saturn.} Univ.\null\ of Arizona
  Press, Tucson. p.\null\ 463                   

\ref Ezer, D., Cameron, A.G.W. (1966) In: {\it Stellar Evolution.}
  Plenum Press, New York.  p.\null\ 203

\ref Fechtig, H., Gr\"un, E., Morfill, G. (1979) {\it Planet.\null\ 
  Space Sci.}\null\ {\bf 27}, 511

\ref Festou, M.C., Atreya, S.K., Donahue, T.M., et al. (1981) 
  {\it J.\null\ Geophys.\null\ Res.}\null\ {\bf 86}, 5715 

\ref Fishman, G.J., Harmon, B.A., Gregory, J.C., et al. 
  (1991) {\it Nature} {\bf 349}, 678

\ref Foster, P.N., Boss, A.P. (1996) {\it ApJ} {\bf 467}, 784
 
\ref Fowler, W.A., Greenstein, J.L., Hoyle, F. (1962a) 
  {\it Geophys.\null\ J.}\null\ {\bf 6}, 148

\ref Fowler, W.A., Caughlan, G.R., Zimmerman, B.A. 
  (1962b) {\it Ann.\null\ Revs.\null\ A\&A} {\bf 5}, 525

\ref Fried, J.W. (1978) {\it A\&A} {\bf 68}, 259

\ref Garrett, H.B. (1981) {\it Rev.\null\ Geophys.\null\ Space 
  Phys.}\null\ {\bf 19}, 577

\ref Glanz, J. (1996) {\it Science} {\bf 272}, 1738

\ref Golub, L., Krieger, A.S., Silk, J.K., Timothy, A.F., 
  Vaiana, G.S. (1974) {\it ApJ} {\bf 189}, L93

\ref Grebowsky, J.M., Pharo III, M.W. (1985) {\it Plan.\null\ 
  Space Sci.}\null\ {\bf 33}, 807

\ref Gregory, J.C., Albrecht, A., Herzog, G., et al. (1993)
  Cosmogenic Radionuclides on LDEF: An Unexpected $\rm^{10}Be$ 
  Result. In: LDEF - 69 Months in Space, {\it Proceedings of the 
  Second Post-Retrieval Symposium.} NASA {\bf CP-3194}, Washington.  
  p.\null\ 231

\ref Gr\"un, E., Pailer, N., Fechtig, H., Kissel, J. (1980) 
  {\it Planet.\null\ Space Sci.}\null\ {\bf 28}, 333

\ref Gr\"un, E., Baguhl, M., Fechtig, H., et al. (1992) 
  {\it Geophys.\null\ Res.\null\ Lett.}\null\ {\bf 19}, 1311

\ref Gr\"un, E., Zook, H.A., Baguhl, M., et al. (1993) {\it 
  Nature} {\bf 362}, 428

\ref Gr\"un, E., Hamilton, D.P., Riemann, R., et al. (1996) {\it Science} 
  {\bf 274}, 399

\ref Gurnett, D.A., Kurth, W.S., Allendorf, S.C., Poynter, R.L.
  (1993) {\it Science} {\bf 262}, 199

\ref Hemenway, C.L., Soberman, R.K., Witt, G. (1963) {\it Nature} 
  {\bf 199}, 269

\ref Hicks, T.R., May, B.H., Reay, N.K. (1974) {\it MNRAS} {\bf 
  166}, 439

\ref Hildner, E. (1986) {\it Adv.\null\ Space Res.}\null\ {\bf 
  6}, 297

\ref Hill, F., Stark, P.B., Stebbins, R.T., et al. (1996)
  {\it Science} {\bf 272}, 1292

\ref Hodapp, K.-W., MacQueen, R.M., Hall, D.N.B. (1992) {\it 
  Nature} {\bf 355}, 707

\ref Houminer, Z., Hewish, A. (1974) {\it Planet.\null\ Space 
  Sci.}\null\ {\bf 22}, 1041

\ref Hovestadt, D., Vollmer, O., Gloeckler, G., Fan, C.Y. (1973)
  {\it Phys.\null\ Rev.\null\ Lett.}\null\ {\bf 31}, 650

\ref Howard, R., Fritzov\' a-\v Svestkov\' a, L., \v Svestka, Z. 
  (1979) {\it Solar Phys.}\null\ {\bf 63}, 105

\ref Jager, C.de (1986) {\it Adv.\null\ Space Res.}\null\ {\bf 
  6}, 353

\ref Kalas, P., Jewitt, D. (1997) {\it Nature} {\bf 386}, 52

\ref Kane, T.J., Mui, P.H., Gardner, C.S. (1992) 
  {\it Geophys.\null\ Res.\null\ Lett.}\null\ {\bf 19}, 405

\ref Kessler, D.J. (1970) {\it Meteoroid Environment Model - 1970 
  [Interplanetary and Planetary].} NASA {\bf SP-8038}, Washington.

\ref Kippenhahn, R., Weigert, A. (1994) {\it Stellar Structure and 
  Evolution.}  Springer Verlag, Berlin.

\ref Klecker, B., Hovestadt, D., Gloeckler, G., Fan, C.Y. (1977)
  {\it ApJ} {\bf 212}, 290 

\ref K\"ohler, H. (1969) {\it Dissertation}, G\"ottingen

\ref Krauss, L.M. (1990) {\it Nature} {\bf 348}, 403                

\ref Lagrange, A.-M., Beuzit, J.-L., Mouillet, D. (1996) {\it J. 
  Geophys.\null\ Res.}\null\ {\bf 101}, 14831

\ref Lamy, P., Kuhn, J.R., Lin, H., Koutchmy, S., 
  Smartt, R.N. (1992) {\it Science} {\bf 257}, 1377

\ref Leising, M.D. (1990) Gamma Ray Lines from 
  Classical Novae. In: Durouchoux, P., Prantzos, N.(eds.) 
  {\it Gamma-Ray Line Astrophysics}, Amer.\null\ Inst.\null\ 
  Phys.\null\ {\bf v.\null\ 126}, New York, p.\null\ 173

\ref Libbrecht, K.G., Morrow, C.A. (1991) The Solar Rotation.
  In Cox, A.N., Livingston, W.C., Mathews, M.S. (eds.) {\it Solar 
  Interior and Atmosphere.} Univ.\null\ of Arizona
  Press, Tucson. p.\null\ 479

\ref Lissauer J.J. (1997) {\it Nature} {\bf 386}, 18

\ref Lisse, C.M., Denneri, K., Englhauser, J., et al. (1996) {\it 
  Science} {\bf 274}, 205

\ref McDonald, F.B., Teegarden, B.J., Trainor, J.H., Webber, 
  W.R. (1974) {\it ApJ} {\bf 187}, L105

\ref McKinley, D.W.R. (1961) {\it Meteor Science and 
  Engineering.} McGraw-Hill, New York   

\ref McNutt Jr., R.L. (1995) {\it Science} {\bf 270}, 1635

\ref Mende, S.B., Swenson, G.R., Miller, K.L. (1985) 
  {\it J.\null\ Geophys.\null\ Res.}\null\ {\bf 90}, 6667

\ref Michels, D.J., Sheeley Jr., N.R., Howard, R.A., Koomen, M.J., 
  1982, {\it Science} {\bf 215}, 1097

\ref Mikheyev, S.P., Smirnov, A.Yu. (1986) {\it Nuovo Cimento} 
  {\bf 9C}, 17

\ref Oakley, D.S., Snodgrass, H.B., Ulrich, R.K., VanDeKop, T.L., 
  (1994) {\it ApJ} {\bf 437}, L63

\ref Olsson-Steel, D., Elford, W.G. (1987)
  {\it J.\null\ Atmos.\null\ Terr.\null\ Phys.}\null\ {\bf 49}, 243

\ref \"Opik, E.J. (1950) {\it Irish Astron.\null\ J.}\null\ {\bf 1}, 80

\ref Ouyed, R., Pudritz, R.E., Stone, J.M. (1997) {\it Nature} 
  {\bf 385}, 409

\ref Parker, E. (1978) In: {\it The New Solar Physics}, Am.\null\ 
  Assoc.\null\ Adv.\null\ Sci.\null\ {\bf v.\null\ 17}, Washington, 
  p.\null\ 8

\ref Peterson, A.M., Villard, O.G., Lederbrand, R.L., 
  Gallagher, P.E. (1960) {\it J.\null\ Geophys.\null\ Res.}\null\ 
  {\bf 60}, 497

\ref Raghavan, R.S. (1995) {\it Science} {\bf 267}, 45

\ref Ramaty, R., Lingenfelter, R.E. (1979) {\it Nature} {\bf 
  278}, 127

\ref Rasio, F.A., Ford, E.B. (1996) {\it Science} {\bf 274}, 954 

\ref Ray, T.P., Muxlow, T.W.B., Axon, D.J., Brown, A., Corcoran, D.,
  Dyson, J., Mundt, R. (1997) {\it Nature} {\bf 385}, 415

\ref Rowley, J.K., Cleveland, B.T., Davis Jr., R., 
  1985, In: {\it Solar Neutrinos and Neutrino Astronomy (Homestake, 
  1984)}.  Amer.\null\ Inst.\null\ Phys.\null\ {\bf v.\null\ 
  126}, New York, p.\null\ 1

\ref Rubin, V.C., Ford, W.K. Jr., Thonnard, N. (1980) 
  {\it ApJ} {\bf 238}, 471

\ref R\"udiger, G. (1989) {\it Differential Rotation and Stellar 
  Convection, Sun and Solar Type Stars.} Gordon \& Breach, New 
  York

\ref Russell, H.N., Dugan, R.S., Stewart, J.W. (1927)
  {\it Astronomy II: Astrophysics and Stellar Astronomy.} Ginn \& Co., 
  Boston, p.\null\ 909  

\ref Ryle, M., Hewish, A. (1950) {\it MNRAS} {\bf 110}, 381

\ref Sandlin, G.D., Bartoe, J.-D.F., Bruchkner, G.E., 
  Tousey, R., VanHoosier, M.E. (1986) {\it ApJS} {\bf 61}, 801

\ref Sargent, A.I., Beckwith, S.V.W. (1993) {\it Phys.\null\ 
  Today} {\bf 46}, 22

\ref Schaeffer, O.A., Zahringer, J. (1962) 
  {\it Phys.\null\ Rev.\null\ Lett.}\null\ {\bf 8}, 389

\ref Schwarzschild, M. (1958) {\it Structure and Evolution 
  of the Stars.} Princeton Univ.\null\ Press, Princeton  

\ref Seiff, A., Kirk, D.B., Knight, T.C.D., et al. (1997) {\it Science} 
  {\bf 276}, 102

\ref Sekanina, Z., Larson, S.M. (1986) {\it Nature} {\bf 321}, 357

\ref Silverman, S.M. (1983) {\it J.\null\ Geophys.\null\ 
  Res.}\null\ {\bf 88}, 8123

\ref Smith, B.A., Soderblom, L.A., Beebe, R., et al. 
  (1981) {\it Science} {\bf 212}, 163

\ref Smith, B.A., Soderblom, L., Batson, R., et al. (1982) 
  {\it Science} {\bf 215}, 504

\ref Soberman, R.K. (1963) {\it Sci.\null\ Amer.}\null\ {\bf 
  208}, 50

\ref Soberman, R.K., Della Lucca, L. (1963) {\it Smithsonian 
  Contr.\null\ Ap.}\null\ {\bf 7}, 85 

\ref Starrfield, S. (1989) Thermonuclear Processes and the 
  Classical Nova Outburst. In: Bode, M. (ed.) {\it Classical Novae.} 
  Wiley, New York, p.\null\ 39

\ref Starrfield, S., Truran, J.W., Sparks, W.M., Arnould, 
  M. (1978) {\it ApJ} {\bf 222}, 600

\ref Tappin, S.J.,  Hewish, A., Gapper, G.R. (1983)
  {\it Planet.\null\ Space Sci.}\null\ {\bf 31}, 1171

\ref Terekhov, O.V., Syunyaev, R.A., Kuznetsov, A.V., et 
  al. (1993) {\it Astron.\null\ Lett.}\null\ {\bf 19}, 65

\ref Thomas, R. N. (1993) Perspective. In: Hack, M., la Douz, C. 
  (eds.) {\it Cataclysmic Variables and Related 
  Objects.} NASA {\bf SP-507}, Washington, p.\null\ xiii

\ref Underhill, A.B. (1955) {\it J.\null\ Roy.\null\ Astron.\null\ Soc.\null\ 
  Can.}\null\ {\bf 49}, 27

\ref Vogt, H. (1926) {\it Astron.\null\ Nachr.}\null\ {\bf 226}, 301 

\ref Warwick, J.W., Evans, D.R., Peltzer, G.R., et al.
  (1989) {\it Science} {\bf 246}, 1498

\ref Webber, W.R., Lockwood, J.A. (1993) {\it J.\null\ Geophys.\null\ 
  Res.}\null\ {\bf 98}, 7821

\ref Wilson, R. (1955) {\it Observatory} {\bf 75}, 222 

\ref Woodard, M.F., Libbrecht, K.G. (1993) {\it Science} {\bf 
  260}, 1778

\ref Zahn, U.von, Hansen, T.L. (1988) {\it J.\null\ Atmos.\null\ 
  Terr.\null\ Phys.}\null\ {\bf 50}, 93

\ref Zwicky, F. (1937) {\it ApJ} {\bf 86}, 217

\end